\documentclass[journal,comsoc]{IEEEtran}

\usepackage[T1]{fontenc}
\usepackage{amssymb, subfigure, setspace, multirow}
\usepackage{verbatim}
\usepackage{authblk}
\usepackage{amsmath}    	
\usepackage[cmintegrals]{newtxmath}
\usepackage{hhline}     		
\usepackage{cite}
\usepackage{algorithmic}
\usepackage{algorithm}
\usepackage{verbatim}
\usepackage{array}
\usepackage{url}
\usepackage{bm}
\usepackage{mathrsfs}
\usepackage{amsfonts}
\usepackage{graphicx}
\usepackage{textcomp}
\usepackage{xcolor}
\usepackage{microtype}
\usepackage{gensymb}

\def\BibTeX{{\rm B\kern-.05em{\sc i\kern-.025em b}\kern-.08em
    T\kern-.1667em\lower.7ex\hbox{E}\kern-.125emX}}

\def\squareforqed{\hbox{\rlap{$\sqcap$}$\sqcup$}}
\def\qed{\ifmmode\squareforqed\else{\unskip\nobreak\hfil
\penalty50\hskip1em\null\nobreak\hfil\squareforqed
\parfillskip=0pt\finalhyphendemerits=0\endgraf}\fi}

\makeatletter
\def\url@leostyle{%
  \@ifundefined{selectfont}{\def\UrlFont{\sf}}{\def\UrlFont{\small\ttfamily}}}
\makeatother
\newcommand{\ie}{{\em i.e.}}
\newcommand{\eg}{{\em e.g.}}

\def\spth{\textsuperscript{th}}

\DeclareMathOperator*{\argmin}{argmin} 

\begin{document}
\title{Robot-assisted Backscatter Localization for IoT Applications}

\author{Shengkai~Zhang,~\IEEEmembership{Student~Member,~IEEE, }
		Wei~Wang,~\IEEEmembership{Senior~Member,~IEEE, }
		Sheyang~Tang, Shi~Jin,~\IEEEmembership{Member,~IEEE, }
        and~Tao~Jiang,~\IEEEmembership{Fellow,~IEEE}
 \thanks{Part of this work has been presented at IEEE GLOBECOM \cite{zhang2019localizing}.}
 \thanks{This work was supported in part by the National Key R\&D Program of China under Grant 2019YFB180003400, Young Elite Scientists Sponsorship Program by CAST under Grant 2018QNRC001, National Science Foundation of China with Grant 91738202.}
\thanks{S. Zhang, W. Wang, S. Tang and T. Jiang are with the School of Electronic Information and Communications, Huazhong University of Science and Technology, Wuhan 430074, China. (E-mail: \{szhangk, weiwangw, sheyangtang, taojiang\}@hust.edu.cn).}
\thanks{S. Jin is with the School of Information Science and Engineering, Southeast University, Nanjing 211189, China. (E-mail: jinshi@seu.edu.cn).}
}

\maketitle

\begin{abstract}

Recent years have witnessed the rapid proliferation of backscatter technologies that realize the ubiquitous and long-term connectivity to empower smart cities and smart homes. Localizing such backscatter tags is crucial for IoT-based smart applications. However, current backscatter localization systems require prior knowledge of the site, either a map or landmarks with known positions, which is laborious for deployment. To empower universal localization service, this paper presents Rover, an indoor localization system that localizes multiple backscatter tags without any start-up cost using a robot equipped with inertial sensors. Rover runs in a joint optimization framework, fusing measurements from backscattered WiFi signals and inertial sensors to simultaneously estimate the locations of both the robot and the connected tags. Our design addresses practical issues including interference among multiple tags, real-time processing, as well as the data marginalization problem in dealing with degenerated motions. We prototype Rover using off-the-shelf WiFi chips and customized backscatter tags. Our experiments show that Rover achieves localization accuracies of $39.3$ cm for the robot and $74.6$ cm for the tags.

\end{abstract}


\IEEEpeerreviewmaketitle

\section{Introduction}
\label{sec:intro}
\IEEEPARstart{I}{n} the last few years, the rapid innovations of small-footprint and low-power backscatters in both end-to-end communication~\cite{kellogg2014wi, peng2018plora, xu2018backscatter, bharadia2015backfi, amato2018rfid, guo2018design} and large-scale networking~\cite{hessar2019netscatter, xia2019ftrack, liu2018backscatter} have been driving the realization of the universal Internet-of-Things (IoT) deployment. Their designs enable concurrent communications among a large number of IoT devices and low-power communications that avoid the inconvenience of changing the battery. Localizing such universal backscatters is crucial for ubiquitous sensing and smart services in both domestic and industrial fields. For example, the location of a dog chew toy can be monitored for the convenience of home cleaning and the RF penetration capability from backscatter tags enables the item tracking in highly cluttered settings for industrial production, \eg, tracking an item from under a pile~\cite{luo20193d} to help a robot pick it.
 
To date, the fundamental challenge of backscatter localization has been its limited communication range due to the low-power signals, which turns out that the localizability is limited in a small region, \eg, a room~\cite{kotaru2017localizing}, or requires dense landmark deployment~\cite{luo20193d} that increases the start-up cost. To overcome this challenge, LoRa's high sensitivity holds the opportunity to enable long-range sensing with extremely low-power signals~\cite{chen2019widesee} and the superior mobility of drones~\cite{ma2017drone} breaks the spatial limitation by approaching the tags and magnify the backscattered signals. Unfortunately, these works either require dedicated RF sources which are not ubiquitously available~\cite{chen2019widesee}, or need additional sensing modalities to obtain the location of the vehicle first~\cite{ma2017drone}.


Ideally, we desire a system that supports the IoT localization with low-power signals that extends the tracking demand from smartphones and wearables to universal objects, such as wallets, keys, and lost items: the system should be power-on-and-go that works without any effort for start-up, \eg, site survey or landmark position setup, and it should be ubiquitously available and lightweight that works with low-cost sensor suite upon existing infrastructure for rapid deployment. 

In this paper, we propose Rover, a power-on-and-go backscatter localization system that works with existing ubiquitous infrastructure, \ie, commodity WiFi, and supports instant deployment with zero start-up cost. The CSI of multiple subcarriers of WiFi packets holds the fine-grained localizability~\cite{wang2017csi}. It is a self-contained system that runs on a robot equipped with WiFi chips and an inertial measurement unit (IMU). Rover simultaneously localizes the robot and the backscatter tags that communicate with the robot. It needs to rove in the work space to connect to more tags for localization. Rover addresses the range challenge with the mobility of a robot, who approaches the tags and localizes them by a simultaneous-localization-and-mapping (SLAM) solution. It leverages spatially different observations to construct multi-view constraints for localizing the tags and the robot.

Despite the advantages of Rover, there are three significant challenges. 
{\em First}, without knowing the positions of a robot or any tag, enough translations and angle-of-arrivals (AoAs)\footnote{Measuring AoA can be achieved by off-the-shelf devices using narrowband signals while measuring time-of-flight (ToF) requires specialized hardware or firmware modifications to produce wideband signals~\cite{vasisht2016decimeter}.} of a tag to the robot at different positions are required to satisfy the requirement of triangulation. A straightforward method to obtain the translation is integrating the accelerations measured by IMU. However, the integration operation will lead to a temporal drift of results due to the inherent sensor noise~\cite{he2017pervasive}. To address this issue, we exploit the drift-free localizability via WiFi AoAs by triangulation. We correct the IMU drift by proposing an AoA-IMU SLAM system that jointly optimizes locations of the robot and the connected backscatter tags with WiFi AoAs and IMU measurements. 

{\em Second}, the optimization framework takes the measurements from WiFi and IMU at different locations and our goal is to find a configuration of such locations that best fit all these measurement constraints. In principle, taking more measurements over the robot's trajectory into account for the optimization would provide more accurate results. However, this also incurs more complex computations and delays the localization, making the robot unable to navigate itself when moving. To bound the computation complexity for real-time processing while achieving high accuracy, Rover employs a sliding window based formulation for the SLAM problem and derives the solution in a graph-based optimization framework. 

{\em Third}, the sliding window based formulation for the SLAM problem involves a marginalization operation that removes old states in the window when obtaining new observations. A simple data marginalization scheme is first-in-first-out (FIFO). However, FIFO cannot handle degenerate motions, \eg, being stationary or moving at a constant velocity, as in this case the data in the window cannot recover the metric scale of environments. We propose a flexible marginalization scheme to address this issue.

\noindent {\bf Results}. We prototype Rover on the programmable robot, iRobot Create 2, equipped with an IMU and an Intel Next Unit of Computing (NUC) that installs an Intel 5300 wireless NIC attached with three antennas. We use the 802.11n Channel State Information (CSI) tool~\cite{halperin2011tool} to obtain wireless channel information for AoA estimation. To implement the backscatter tag, we use the hardware provided by HitchHike~\cite{zhang2016hitchhike} and reprogram its FPGA with our Rover firmware. The experiments are conducted with four backscatters deployed in a conference room of our laboratory to validate individual system modules as well as the overall performance. The results show that Rover is capable of handling the robot's degenerated motions and achieves localization accuracies of $74.6$ cm for the backscatter tag and $39.3$ cm for the robot over a trajectory of $41.96$ m.

\noindent {\bf Contributions}. Rover is the first backscatter localization system that works with a single robot using commodity WiFi without any prior knowledge of work space. Rover leverages the localizability of WiFi signals to correct the IMU drift. {\em First}, we propose an AoA-IMU SLAM system that jointly optimizes the locations of the robot and the connected backscatter tags subject to measurement constraints of IMU and WiFi. {\em Second}, we employ a sliding window based formulation for the SLAM problem to achieve real-time processing. {\em Third}, we devise a flexible marginalization scheme for the sliding window operation to handle degenerated motions. We implement Rover on commodity devices and experimentally validate the system in indoor environments.

The rest of this paper is divided into four parts. Section~\ref{sec:background} introduces the interference avoidance scheme and the AoA estimation technique. Section~\ref{sec:design} presents the sliding window based AoA-IMU SLAM system as well as the flexible marginalization scheme. We show the results of system evaluations in Section~\ref{sec:evaluation} and summarize the related works in Section~\ref{sec:related}. Finally, Section~\ref{sec:conclusion} concludes this work and points out possible directions for future improvements.

\section{Backscatter AoA Estimation}
\label{sec:background}
\begin{figure}[t!]
  \centering
  \includegraphics[width=2.5in]{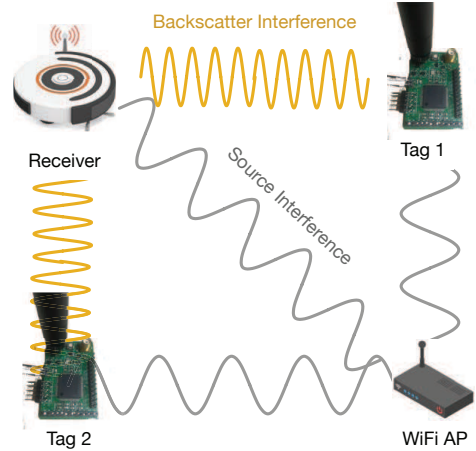}
  \caption{The interference between the backscattered signals (yellow wave) from tags and the excitation signal (grey wave) from a transmitter.}
  \label{fig:background_a}
\end{figure}

Differing from conventional WiFi localization systems that the target device responds to a WiFi receiver through an active WiFi radio, backscatter localization systems arise two problems: 1) The WiFi receiver should be able to receive and decode the backscattered low-power signals from tags; 2) Multiple tags that concurrently backscatter signals can interfere with each other or with the WiFi transmitter as shown in Fig.~\ref{fig:background_a}. The key to address the first problem is to make sure the tag reflects the preamble of the excitation WiFi packet. We implement an envelop detector to detect the starting point of a WiFi packet and backscatter a decodable WiFi packet as proposed in~\cite{kotaru2017localizing}. To resolve the interference, frequency shifting~\cite{zhang2016hitchhike} is effective that we can build a tag that shifts the WiFi signal by a particular frequency to another channel and then backscatters the frequency-shifted signal. Multiple tags need to shift into different channels. While the number of WiFi channels is limited, we propose an interference avoidance mechanism that allows Rover to work in a scalable battery-free network. 

\subsection{Interference Avoidance}

To avoid the interference from a WiFi transmitter, a straightforward solution is to move the backscattered signal to another channel that does not overlap with the channel where the excitation WiFi signal is sent. We achieve this by toggling its RF transistor at a higher speed~\cite{zhang2016hitchhike}, \eg, $20$ MHz. Then, the backscattered signal will be moved to a channel that is $20$ MHz away from the channel where the excitation WiFi signal stays. Configuring the receiver to work on the channel of the backscattered signal will address the interference from the excitation signal.

To avoid the interference from multiple tags that concurrently backscatter signals, the above idea can be extended to assign different available WiFi channels to each of the tags by toggling their RF transistors to different speeds. In this way, the tags can be distinguished by their allocated WiFi channels. Each channel corresponds to a particular tag. To receive packets from all tags, the receiver in our system needs to be capable of sweeping all WiFi channels except the channel of the excitation signal. We achieve this by implementing a frequency band sweeping protocol~\cite{vasisht2016decimeter} in the iwlwifi driver of Intel 5300 NIC. Since the number of non-overlapped WiFi bands is limited, Rover can only simultaneously localize a limited number of tags. To maximize the ability of simultaneous localization, it is vital to choose the channel of the excitation signal. 

Suppose a tag uses a frequency $f_b$ square wave signal to control the on-off frequency of the RF switch. $f_c$ is the carrier center frequency of the 802.11n excitation signal. Let $\omega_b = 2\pi f_b$, $\omega_c = 2\pi f_c$, and $\alpha_{\text{base}}(t)$ denotes a baseband waveform. The square wave can be formulated as $S_{\text{tag}} = \frac{4}{\pi}\sum_{n=1}^{\infty}\frac{\sin\left[(2n-1)\omega_b t\right]}{2n-1}$. Hence, the backscattered signal $\beta(t)$ can be written as,
\begin{equation}
	\beta(t) = \alpha_{\text{base}}(t)e^{j\omega_c t}S_{\text{tag}}(t).
\end{equation}
Let $F_{\text{base}}(\omega)$ and $F(\omega)$ be the Fourier transform of $\alpha_{\text{base}}(t)$ and $\beta(t)$ respectively. We have 
\begin{equation}
	\begin{aligned}
		F(\omega) =  & \sum_{n=1}^{\infty}\frac{2j}{\pi (2n-1)}\left( F_{\text{base}} \left( \omega - \omega_c + (2n-1)\omega_b \right) - \right. \\
		& \left. F_{\text{base}} \left( \omega - \omega_c - (2n-1)\omega_b \right) \right).
	\end{aligned}
\end{equation}
This indicates that the frequency-shifted backscattered signal can be received at two bands, $f_c \pm f_b$, causing sideband interference to other channels. 
Based on this, Rover chooses the most side channels in the band, \ie, channel $165$ or $36$, to transmit the the excitation signal in order to avoid the sideband interference.

\subsection{AoA Estimation}

\begin{figure}[t!]
  \centering
  \includegraphics[width=3in]{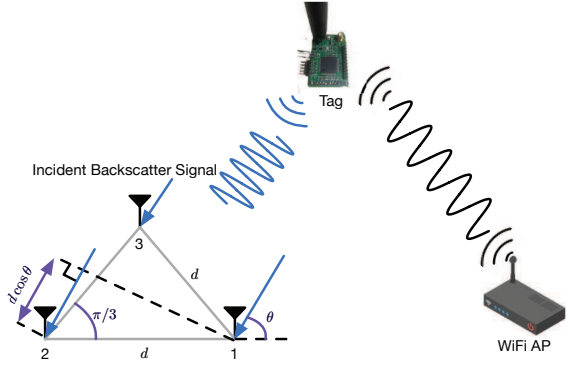}
  \caption{A uniform circular array that consists of three antennas in our system: The signal with AoA $\theta$ travels an additional distance of $d\cos\left(\theta + \frac{\pi}{3}\right)$ to the third antenna and $d\cos(\theta)$ to the second array in the array compared to the first antenna.}
  \label{fig:background_b}
\end{figure}

So far, we have addressed the interference problem. A WiFi receiver can receive the backscatter packets from tags. In this section, we describe how Rover estimates the AoAs of backscatter tags to the receiver, which is amounted on a robot, leveraging the CSI of received packets. The AoA estimation technique for low-power backscatter tags was first proposed in~\cite{kotaru2017localizing}. Here we extend it to work with a circular antenna array with uniform spacing $d$ that can measure AoAs in $[0, 360]$ degrees as shown in Fig.~\ref{fig:background_b}.

\begin{figure}[t!]
  \centering
  \includegraphics[width=3in]{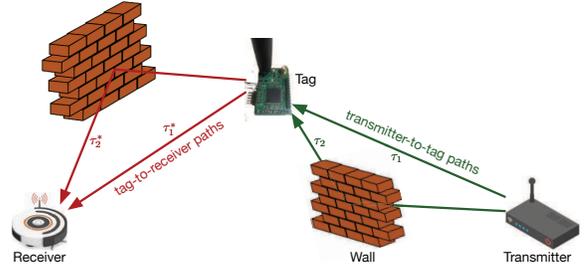}
  \caption{The received signal traverses two physical paths where $\tau_j$ denotes the time of flight (ToF) of $j$\spth path on the transmitter-to-tag link and $\tau^*_i$ denotes the traversing time of $i$\spth path on the tag-to-receiver link.}
  \label{fig:paths}
\end{figure}

Localizing backscatter tags involves two physical paths, transmitter-to-tag path and tag-to-receiver path, as shown in Fig.~\ref{fig:paths}. Thus, the received CSI depends on the locations of backscattered tags and the access point (AP). We combine $j$\spth path on the transmitter-to-tag link with $i$\spth path on the tag-to-receiver link to form a virtual path between the excitation source (AP) and the receiver at the robot. The virtual path has a ToF of $\widehat{\tau}_k = \tau_j + \tau_i^*$ where $\tau_i^*$ ($\tau_j$) denotes the ToF of the signal along a path on the tag-to-receiver (transmitter-to-tag) link, the AoA of the virtual path $\widehat{\theta}_k = \theta_i^*$ where $\theta_i^*$ is the AoA of $i$\spth path on the tag-to-receiver link, and the corresponding complex attenuation of $\widehat{\gamma}_k = \gamma_j \gamma_i^*$ where $\gamma_i^*$ and $\gamma_j$ denote the complex attenuation along $i$\spth path on the tag-to-receiver link and $j$\spth path on the transmitter-to-tag link, respectively. The overall signal obtained at the three antennas for $n$\spth subcarrier can be written as
\begin{equation}
	\begin{aligned}
		H_{n, m} & = \sum_{k=1}^{L_{\text{tx}} L_{\text{tag}} }\widehat{\gamma}_k e^{-j2\pi\left(\widehat{\tau}_k(n-1)f_\delta + (m-1)d \cos\widehat{\theta}_k/\lambda\right)}, m = 1, 2	\\
		H_{n, 3} & = \sum_{k=1}^{L_{\text{tx}} L_{\text{tag}} }\widehat{\gamma}_k e^{-j2\pi\left(\widehat{\tau}_k(n-1)f_\delta + d \cos\left(\widehat{\theta}_k+\frac{\pi}{3}\right)/\lambda\right)},
	\end{aligned}
\end{equation}
where $L_{\text{tag}}$ is the number of paths on the tag-to-receiver link, $L_{\text{tx}}$ is the number of paths on the transmitter-to-tag link, $f_\delta$ is the frequency gap between two consecutive subcarriers. This overall signal is reported as CSI corresponding to the particular subcarrier and antenna.

The signal model is a standard form to apply a joint AoA-ToF estimation technique~\cite{kotaru2015spotfi}. The insight of this technique is that multiple subcarriers of an OFDM signal encode ToF information. By smoothing the subcarriers represented in the CSI matrix, it allows a super-resolution AoA estimation with a small antenna array, \eg, a three-antenna array available for Intel 5300 NIC, jointly estimating AoAs and ToFs\footnote{This ToF cannot correctly infer the traveling distance of a propagation path due to its poor distance resolution from the narrowband signal.} of all paths. The AoA of the path with the smallest ToF is the is the direct-path AoA of a tag to the receiver. 

Unfortunately, the obtained AoA can be corrupted by the heading direction of the robot. The robot has three degrees of freedom, including 2D position and the heading direction. It can rotate and change its heading direction while moving, \eg, turning at a corner of the room. Thus, the onboard antenna array will be turning with the robot together. The system can no longer use the measured AoA to localize the robot via triangulation because the AoA not only encodes the geometric constraint of translations but also manifests the rotation (refer to Fig.~\ref{fig:principle} for details). Therefore, we need to correct the rotation from the measured AoA and recover the angle that only relates to the translation.

\begin{figure}[t!]
  \centering
  \includegraphics[width=3.5in]{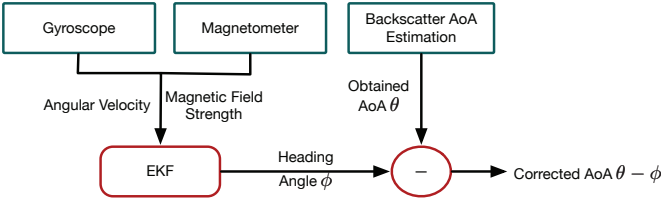}
  \caption{The workflow of AoA correction. We first use the extended Kalman filter to fuse the magnetic field strength with the angular velocity to correct the drift of gyroscope. Then we can trivially correct the AoA.}
  \label{fig:aoa_correction}
\end{figure}

Fig.~\ref{fig:aoa_correction} shows the workflow of AoA correction. Basically, we leverage the IMU to estimate the robot's heading direction in angle $\phi$, assuming that the initial heading direction is angle $0\degree$. The gyroscope in IMU provides raw measurements of angular velocity. However, it is well-known that simply integrating them to obtain the heading will result in error accumulation. We correct the heading by using a magnetometer onboard that provides a reference direction represented by the magnetic field strength. Note that in indoor venues, the reference direction is not the earth's North due to the magnetic interference from surrounding electronic devices. Nevertheless, the reference direction is stable in few hours so that it is eligible to correct the drift during the trajectory. Since the rotation estimation is non-linear, we employ the extended Kalman filter (EKF) to determine the heading by fusing the measurements from gyroscope and magnetometer~\cite{sabatini2006quaternion}. Finally, we correct the AoA by subtracting the heading angle $\phi$ from the obtained AoA $\theta$.

At this stage, we obtain the corrected direct-path AoAs of multiple tags to the receiver. Next, we fuse them with the IMU measurements to localize the tags and the robot simultaneously.

\section{SLAM with AoAs}
\label{sec:design}
Without knowing the position of any device, how can we localize the target by a single mobile robot? In this section, we answer this question by first describing the design of our AoA-IMU localization system and then elaborating on the sliding window based formulation. Finally, we propose a flexible marginalization scheme for addressing a practical issue of sliding window operations in degenerated motions.

\subsection{AoA-IMU Localization System}

The angle (AoA) can be used to determine a target's location via triangulation. Recall that conventional localization systems usually require a few landmarks with known locations to localize the target. The essence of this requirement is defining the metric scale of environments, \ie, the unit (meter, millimeter, etc.) in measuring distances between objects, to fix the size of triangles. 

\begin{figure}[hp]
  \centering
  \includegraphics[width=3.2in]{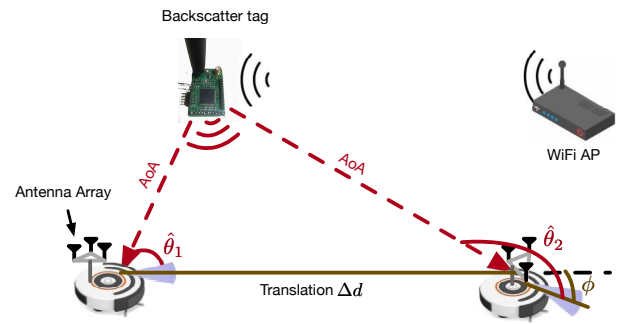}
  \caption{Localization principle: triangulation with the robot's motions. The AP sends WiFi packets to excite the backscatter tag. The receiver on the robot measures the AoAs of the tag to the robot from backscatter signals and the onboard IMU measures translation $\Delta d$ to provide the metric scale of environments. $\hat{(\cdot)}$ denotes the measured AoA and $\phi$ is the rotation of the robot since the previous state.}
  \label{fig:principle}
\end{figure}

In Rover, since the location of both the tags and the robot are unknown, the AoA we obtained cannot yield locations with the metric scale of environments. Nevertheless, with the aid of the onboard IMU and the mobility of a robot, we can localize both the connected tags and the robot in that the IMU provides accelerations in unit $m/s^2$, defining the metric scale. With the translations and the AoAs of incident signals at different positions, it forms a fixed triangle. We take one tag as an example illustrated in Fig.~\ref{fig:principle}. As a robot moves, the IMU measures translation $\Delta d$ and the antenna array measures AoAs $\theta_1$ and $\theta_2$ referring to the tag at different positions, one can determine the relative positions of the robot and the tag through triangulation. Note that the measured AoA has to be corrected from the rotation $\phi$ so as to obtain the correct geometric constraint between the tag and the robot.

Obtaining the translation by integrating the accelerations from the IMU is straightforward but suffers from temporal accumulated errors due to the inherent noise~\cite{he2017pervasive}, causing large localization errors once the result severely distorts the triangle in Fig.~\ref{fig:principle}. To address this issue, we develop an AoA-IMU SLAM approach that optimizes the locations of the robot and backscatters subject to measurement constraints with respect to WiFi AoAs and the IMU odometry. 

Roughly speaking, the central idea of SLAM is to obtain a maximum likelihood estimate of both robot positions and environment features (backscatter tags in our system) given observations (AoAs) from the antenna array. Solutions to the SLAM problem can be either filtering-based or graph-based approaches. While filtering-based approaches are considered to be more efficient in computation~\cite{lu2019collaborative}, we choose graph-based approaches that can achieve better performance via repetitively linearizing past robot states and multi-view constraints~\cite{lin2018autonomous}. 

In addition, solving the SLAM problem is a batch process that incorporates multiple observations to produce accurate results. However, it can become unacceptably slow as the size of the environment grows. This delays the location estimates of the robot so that the robot loses its own navigation capability, being unable to move along the desired trajectory. To let our system run in real-time, we employ an incremental update method to speed up the computation. We formulate a sliding window based model that only keeps a limited amount of AoAs and corresponding robot hidden {\em states}, \eg, the positions of the robot at different timestamps in the workspace, to bound the computation complexity. 

\subsection{System Overview}
\label{subsec:overview}
\begin{figure*}[t!]
  \centering
  \includegraphics[width=6in]{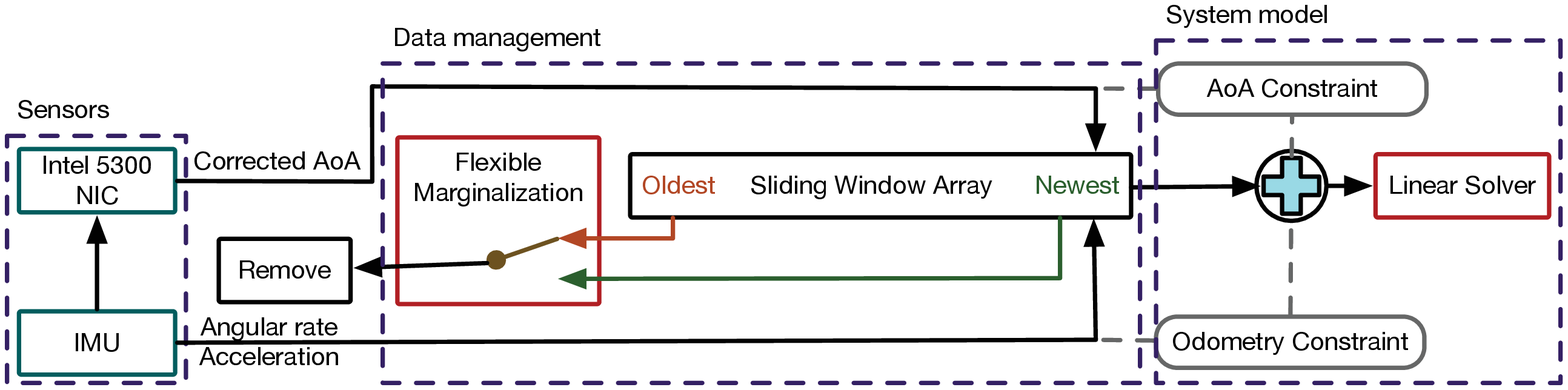}
  \caption{The overview of the SLAM-based system.}
  \label{fig:slam}
\end{figure*}

Upon the introduction of the AoA-IMU localization principle, we give an overview of our system as shown in Fig.~\ref{fig:slam}. Basically, Rover uses two sensors: the Intel 5300 WiFi NIC as an exteroceptive sensor that observes the AoAs with respect to backscatter tags and the IMU as an interoceptive sensor that observes the dynamics of the robot. The sensor data are buffered in a sliding window array for bounding the computation complexity. Then the SLAM-based system model takes the data to solve out the locations of the tags as well as the robot.

Intuitively, the AoA observed by the NIC imposes a geometrical constraint that imply the relative locations of the robot and the connected tags. Note that only the corrected AoA (refer to Fig.~\ref{fig:aoa_correction}) manifests the constraint. To localize the tags, we need to move the robot and capture the dynamics of the robot by IMU. The IMU provides the odometry constraint that indicates the locations of the robot by integrating angular rates and accelerations. Although it is well known that such an integration suffers from a temporal drift, the drift can be corrected by combining the AoA constraint (refer to Section~\ref{subsec:formulation}).

To limit the states and observations in the sliding window, we need to marginalize the data in the window when new observations come. A vanilla option of data marginalization is first-in-first-out (FIFO) that marginalizes out the oldest state and its corresponding measurements. This however cannot handle degenerate motions, \eg, being stationary or moving at a constant velocity. Specifically, if a robot stays at a position for a moment, the measurements keep updating and rendering the sliding window so that the data in the window are all related to the same position. This cannot correctly recover the metric scale by triangulation because the translation $\Delta d$ (Fig.~\ref{fig:principle}) almost diminishes. If a robot moves at a constant velocity, the translation cannot be correctly measured by the IMU due to zero acceleration. Therefore, we propose a flexible marginalization scheme to properly manage the data in the sliding window (refer to Section~\ref{subsec:marginalization}).

\subsection{Sliding Window Formulation}
\label{subsec:formulation}
Table~\ref{table:notation} describes the mathematical notation used in the SLAM algorithm, listed in the order they appear in the text.

\begin{table*}[h]
	\centering  
	\caption{Mathematical notation.}	
	\label{table:notation}  
		\begin{tabular}{{c}l*{1}{c}}  
		Symbol	&	Description	\\
		\hline  		
		$\hat{(\cdot)}$						&	the quantity that can be measured by sensors, \ie, antenna array and IMU	\\
		$\bm{\mu}_i$						&	the hidden state at discrete timestamp $i$ \\
		$\mathbf{o}_{i}^j$				&	the geometric observation from the AoA of backscatter $j$ at timestamp $i$ \\
		$\mathbf{b}_j$						&	the position of backscatter $j$ \\
		$\mathbf{u}_{i+1}^i$			&	the relative translation between two robot states $\bm{\mu}_i$ and $\bm{\mu}_{i+1}$ \\
		$\bm{\mathcal{S}}$				&	the state vector in the sliding window\\
		$\mathcal{A}$						&	the set of AoA measurements between all tags and the robot in the window \\
		$\mathcal{I}$						&	the set of all inertial measurements in the window \\
		$\mathbf{Q}_{i}^{j}$				&	the information matrix of the AoA constraint	\\
		$\mathbf{\Omega}_{i}^j$		&	the AoA covariance matrix	\\
		$\theta$								&	the corrected AoA	\\
		$\mathbf{r}_i^j$					&	the direction vector referred to $i$\spth tag at timestamp $j$	\\
		$d_i^j$									&	the distance between the robot and $i$\spth tag at timestamp $j$	\\
		$\mathbf{n}_i^j$					&	the measurement noise \\
		$\mathbf{P}_{k+1}^{k}$		&	the information matrix of the odometry constraint	\\
		$\bm{\Lambda}_{k+1}^k$	&	the IMU covariance matrix	\\
		$\mathbf{a}_t$						&	the acceleration at current time $t$	\\
		$\bm{\omega}_t$					&	the angular rate at current time $t$	\\
		$\mathbf{R}_t^k$					&	the incremental rotation matrix from time $k$ to current time $t$	\\
		$\mathbf{V}_{k+1}^k$			&	the robot's relative velocity between timestamp $k$ and $k+1$	\\
		$\mathbf{T}_{k+1}^k$			& the robot's relative translation between timestamp $k$ and $k+1$	\\
		$\bm{\nu}_{k}$						&	the velocity at timestamp $k$	\\
		$\textbf{g}$							&	the earth's vertical gravity	\\
		$\Delta t$								&	the time interval between two consecutive measurements	\\
		\hline
	\end{tabular}
\end{table*}

With the AoAs and IMU measurements, we can fuse them to solve the SLAM problem. In this topic, there exists many sensor fusion methods, \eg, EKF, particle filter. However, they usually requires a good initialization, which is very hard to obtain by AoAs due to the lack of metric scale information. Moreover, although filter-based approaches are very efficient in computation as they only estimate the current robot state and the map, the main drawback is that fixing the linearization points early may lead to suboptimal results. Therefore, we employ a graph-based SLAM framework in that 1) it achieves better performance via repetitively linearizing past robot states~\cite{lin2018autonomous}; 2) it is insensitive to the initialization because the multi-view constraint can help recover the initial state.

\begin{figure}[t!]
  \centering
  \includegraphics[width=3in]{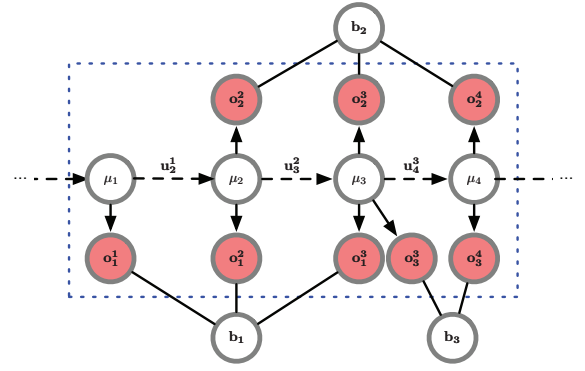}
  \caption{Graph representation for sliding window AoA-SLAM. $\bm{\mu}$ is the hidden state of robot position; $\mathbf{b}$ denotes the hidden state of backscatter position; $\mathbf{o}$ denotes the AoA observation; $\mathbf{u}$ is the odometry captured by IMU. The sliding window represented by the blue dashed box contains four states and their observed AoAs.}
  \label{fig:estimator}
\end{figure}

Fig.~\ref{fig:estimator} shows the graph representation of our SLAM formulation. Let $\bm{\mu}_i$ denote the hidden state at discrete timestamp $i$. At each timestamp, the robot observes a set of AoAs from multiple backscatters. $\mathbf{o}_{i}^j$ is the geometric observation from the AoA of backscatter $j$ at timestamp $i$ and $\mathbf{b}_j$  denotes the position of backscatter $j$. The relative translation between two robot states $\bm{\mu}_i$ and $\bm{\mu}_{i+1}$ is captured by an odometry edge $\mathbf{u}_{i+1}^i$, which can be obtained by IMU preintegration techniques~\cite{forster2015rss}. 

We define the state vector in the sliding window that merges the hidden variables of robot and backscatter together, 
\begin{equation}
  		\bm{\mathcal{S}} 	= [\bm{\mu}_0, \bm{\mu}_1, \dotsc, \bm{\mu}_{n-1}, \mathbf{b}_0, \mathbf{b}_1, \dotsc, \mathbf{b}_{m-1}]^\top,          
\label{eqn:state}
\end{equation}
\noindent where the initial position $\bm{\mu}_0 = [0, 0, 0]$. All these variables refer to the world frame, which is related to the real world where the gravity is vertical. $n$ is the number of robot's state in the sliding window, $m$ denotes the number of observed backscatter tags, and $\mathbf{b}_i$ is the position of tag $i$ in the world frame. At this stage, we have constructed the graph from the AoA observations and the IMU odometry. Next step we seek to find the configuration of the positions of the robot and tags that best satisfies the constraints, \ie, the edges of the graph.

Since our system only involves translations, parameters in $\bm{\mathcal{S}}$ are in Euclidean space. We can formulate the problem as a linear problem and the optimal state sequence $\bm{\mathcal{S}}^*$ in the sliding window can be estimated by solving:
\begin{equation}
		\bm{\mathcal{S}}^* = \argmin_{\bm{\mathcal{S}}} \Big\{\overbrace{\mathbf{A}(\bm{\mathcal{S}})}^\text{AoA constraint} + \overbrace{\mathbf{D}(\bm{\mathcal{S}})}^\text{odometry constraint} \Big\},
  	\label{eqn:cost}
\end{equation}
where 
\begin{equation}
	\begin{aligned}
		\mathbf{A}(\bm{\mathcal{S}}) &=  \sum_{(i, j)\in\mathcal{A}}\left\|\hat{\mathbf{o}}_{i}^{j} - \mathbf{Q}_{i}^{j}\bm{\mathcal{S}}\right\|^2_{\bm{\Omega}_{i}^j} \\
		\mathbf{D}(\bm{\mathcal{S}}) & = \sum_{k\in\mathcal{I}}\left\|\hat{\mathbf{u}}_{k+1}^{k} - \mathbf{P}_{k+1}^{k}\bm{\mathcal{S}}\right\|^2_{\bm{\Lambda}_{k+1}^k}.
	\end{aligned}
  	\label{eqn:cost_specific}
\end{equation}
\noindent $\mathcal{A}$ denotes the set of AoA measurements between all tags and the robot in the window. $\mathcal{I}$ denotes the set of all inertial measurements in the window. The constraints are the sum of the Mahalanobis norm of their measurement errors. Specifically, the AoA constraint is 
\begin{equation}
	\left\|\hat{\mathbf{o}}_{i}^{j} - \mathbf{Q}_{i}^{j}\bm{\mathcal{S}}\right\|^2_{\bm{\Omega}_{i}^j} = \left(\hat{\mathbf{o}}_{i}^{j} - \mathbf{Q}_{i}^{j}\bm{\mathcal{S}}\right)^\top \left(\bm{\Omega}_i^j\right)^{-1} \left(\hat{\mathbf{o}}_{i}^{j} - \mathbf{Q}_{i}^{j}\bm{\mathcal{S}}\right), 
\end{equation}
 and the odometry constraint is 
 \begin{equation}
 \begin{aligned}
 	& \left\|\hat{\mathbf{u}}_{k+1}^{k} - \mathbf{P}_{k+1}^{k}\bm{\mathcal{S}}\right\|^2_{\bm{\Lambda}_{k+1}^k} = \\ & \left(\hat{\mathbf{u}}_{k+1}^{k} - \mathbf{P}_{k+1}^{k}\bm{\mathcal{S}}\right)^\top \left(\bm{\Lambda}_{k+1}^k\right)^{-1} \left(\hat{\mathbf{u}}_{k+1}^{k} - \mathbf{P}_{k+1}^{k}\bm{\mathcal{S}}\right).
 \end{aligned}
 \end{equation}
To solve this system, the terms of the AoA constraint $\left\{\hat{\mathbf{o}}_{i}^{j}, \mathbf{Q}_{i}^{j}, \mathbf{\Omega}_{i}^j\right\}$ and the odometry constraint $\left\{\hat{\mathbf{u}}_{k+1}^{k}, \mathbf{P}_{k+1}^{k}, \bm{\Lambda}_{k+1}^k\right\}$ need to be defined.

{\bf AoA constraint}. The direction vector $\mathbf{r}_{i}^{j}$ referred to the observed $i$\spth tag at timestamp $j$ can be defined by the AoA $\theta$ as $\mathbf{r}_{i}^{j} = [\cos(\theta), \sin(\theta), 0]^{\top}$. With an unknown distance $d_i^j$, a simple geometric relationship can be expressed as
\begin{equation}
  d_i^{j}\mathbf{r}_{i}^{j} = \mathbf{R}_{0}^{j}\left(\mathbf{b}_{i} - \bm{\mu}_{j}\right),
  \label{eqn:similar_equation}
\end{equation}
\noindent where $\mathbf{b}_{i}$ is the $i$\spth tag's position and $\bm{\mu}_j$ is the robot position at timestamp $j$. Since $\mathbf{r}_{i}^{j}$ should have the same direction as the vector $\mathbf{b}_{i} - \bm{\mu}_{j}$ if there is no measurement noise. The expected observation can be expressed by a cross product operation,
\begin{equation} 
  \hat{\mathbf{o}}_i^{j} = \hat{\mathbf{0}} = \left(\mathbf{R}_{j}^{0}\mathbf{r}_{i}^{j}\right) \times \left(\mathbf{b}_{i} - \bm{\mu}_{j}\right) = \mathbf{Q}_i^{j}\bm{\mathcal{S}} + \mathbf{n}_i^{j},
  \label{eqn:wifi_measurement_model}
\end{equation}
\noindent where $\mathbf{n}_i^{j}$ denotes the noise, assuming that it follows a Gaussian distribution. The AoA covariance $\bm{\Omega}_i^j$ can be pre-measured by statistical methods and updated along the optimization process. Initially, the distance $d_{i}^{j}$ is given by a reasonable guess. Then it will be refined automatically along the sliding window optimization as the positions of the robot and tags are updated. Therefore, the initial guess is insensitive in our system.


Note that we consider the AoA in 2D case for the ease of representation. Our system can be trivially extended to work in 3D case. The circular antenna array we use is capable of measuring azimuth $\theta$ angle and elevation angle $\psi$ for 3D AoA representation. The direction vector becomes $\mathbf{r} = \left[ \cos\theta \sin\psi,\; \sin\theta \sin\psi, \; \cos\psi \right]$. However, we can no longer employ the joint AoA-ToF estimation technique~\cite{kotaru2015spotfi} to obtain the 3D AoA as the joint parameter searching process of this technique will increase the computation complexity exponentially due to the additional parameter, \ie, $\psi$. To reduce the complexity, we employ an additional parameter search instead of the joint search. This is an approximate solution of~\cite{kotaru2015spotfi} that slightly sacrifices the accuracy to significantly save the computation cost. Its computation complexity remains the same as the 2D case. It may occasionally miss the optimal parameter configuration but the overall performance is very close to the optimal solution as proved by~\cite{xie2019md}. Since the 3D extension is incremental to our contribution, we omit the details in this paper.


{\bf Odometry constraint}. Typically, the data rate of IMU is higher than AoA rate. Given two consecutive timestamps $[k, k+1]$ at which the AoAs from multiple tags are received, there have been multiple buffered inertial measurements, which include acceleration $\mathbf{a}_t \in \mathbb{R}^{3}$ and angular rate $\bm{\omega}_t \in \mathbb{R}^{3}$. We can preintegrate them to obtain an overall odometry representation between $\bm{\mu}_k$ and $\bm{\mu}_{k+1}$ as follows:

\begin{equation}
  \begin{aligned}
  	\mathbf{V}_{k+1}^{k}  &= \sum_{t\in[k, k+1]}\mathbf{R}_t^{k}\mathbf{a}_t \Delta t     \\
    \mathbf{T}_{k+1}^{k} &= \sum_{t\in[k, k+1]}\left[\mathbf{V}_{k+1}^{k}\Delta t + \mathbf{R}_t^{k}\mathbf{a}_t\Delta t^2\right], 
  \end{aligned}
  \label{eqn:preimu}
\end{equation}
\noindent where $\mathbf{R}_t^{k} = \sum_{i \in [k, t]}\left[\mathbf{R}_i^{k}\lfloor\bm{\omega}_t\times\rfloor \Delta t\right]$,  $\mathbf{R}_t^{k} \in \text{SO}(3)$. $\lfloor\bm{\omega}_t\times\rfloor$ is the skew-symmetric matrix from $\bm{\omega}_t$, $\Delta t$ the time interval between two consecutive measurements. $\mathbf{R}_t^{k}$ denotes the incremental rotation from time $k$ to current time $t$, which is available through short-term integration of gyroscope measurements. Then, we can write the propagation model of positions as
\begin{equation}
    \bm{\mu}_{k+1}  = \bm{\mu}_{k} + \mathbf{R}_{k}^0\bm{\nu}_{k}\Delta t - \mathbf{R}_{k}^{0}\textbf{g}\Delta t^2/2 + \mathbf{R}_{k}^0\mathbf{T}_{k+1}^{k}, 
  	\label{eqn:linear_update}
\end{equation}
\noindent where $\mathbf{T}_{k+1}^{k}$ can be obtained by Eqn.~\eqref{eqn:preimu}. $\textbf{g} = [0, 0, 9.8]^\top$ is the vertical gravity. Since the robot only moves in a room (assuming a horizontal plane), it is safe to obtain the accelerations that account for motions by directly subtracting the gravity. $\bm{\nu}_k$ denotes the velocity at timestamp $k$. It can be propagated as 
\begin{equation}
	\bm{\nu}_{k+1} = \mathbf{R}_{k}^{k+1}\bm{\nu}_{k} - \mathbf{R}_{k}^{k+1}\textbf{g}\Delta t + \mathbf{R}_{k}^{k+1}\mathbf{V}_{k+1}^{k},
	\label{eqn:velocity_update}
\end{equation}
where $\mathbf{V}_{k+1}^{k}$ is obtained from Eqn.~\eqref{eqn:preimu}. $\mathbf{R}_{k}^{0}$ is the change in rotation since the initial state. We can see that the update equation for the quantity $\bm{\mu}_{k}$ and $\bm{\nu}_{k+1}$ will be linear in Eqn.~\eqref{eqn:linear_update} and Eqn.~\eqref{eqn:velocity_update} if rotation $\mathbf{R}_{k}^{0}$ are provided. This rotation can be obtained by solving a linear system that incorporate the short-term integration of gyroscope measurements. For brevity, we omit the details and refer to the broad literature discussing these ideas~\cite{shen2016initialization}.

Accordingly, Eqn.~\eqref{eqn:linear_update} can be rewritten as a linear function of the state $\bm{\mathcal{S}}$: 
\begin{equation}
  \begin{aligned}
    \hat{\mathbf{u}}_{k+1}^{k} = \hat{\mathbf{T}}_{k+1}^{k}   
    & = \mathbf{R}_{0}^{k}\left(\bm{\mu}_{k+1} - \bm{\mu}_{k}\right) - \bm{\nu}_{k}\Delta t + \textbf{g}\frac{\Delta t^2}{2}  \\
    &= \mathbf{P}_{k+1}^{k}\bm{\mathcal{S}} + \mathbf{n}_{k+1}^{k},
  \end{aligned}
  \label{eqn:imumodel}
\end{equation}
\noindent where $\bm{\nu}_k$ can be updated by Eqn.~\eqref{eqn:velocity_update}, $\mathbf{n}_{k+1}^{k}$ denotes the additive measurement noise. Typically, we assume the additive noise follows a Gaussian distribution. Then the covariance $\bm{\Lambda}_{k+1}^{k}$ can be calculated using the pre-integration technique proposed in~\cite{lupton2012visual}.

At this point, all constraints in Eqn.~\eqref{eqn:cost_specific} are explicitly defined. The information matrices and state vectors in the sliding window can be stacked to construct a large array of linear equations so that the positions of the robot and the tags in the window can be solved altogether. 

However, the robot may undergo some degenerated motions in practical, causing the data marginalization problem as mentioned in Section~\ref{subsec:overview}. We next elaborate on our novel marginalization scheme.

\subsection{Flexible Marginalization}
\label{subsec:marginalization}
The FIFO marginalization scheme works fine when the robot is moving with non-zero acceleration. However, this scheme fails when the robot performs degenerate motions (zero acceleration), \eg, being stationary or moving in a constant velocity. In these cases, the IMU odometry, \ie, the translations between AoAs, cannot be correctly measured, failing to recover the metric scale. Unfortunately, zero acceleration motion is unavoidable in practice for a mobile robot and it must be handled properly.

When being stationary, the FIFO scheme results in that all measurements in the sliding window come from the same position. The translation between two AoAs is unobservable so that we cannot recover the metric scale. Intuitively, the last-in-first-out (LIFO) sliding window scheme can preserve the scale observability. In this case, we only update the position of the robot because LIFO scheme does not keep new AoAs. 

When moving at a constant velocity, the translations between AoAs cannot be correctly measured, making the metric scale still unobservable. For example, if the robot first undergoes generic motions with sufficient accelerations ($\bm{\mu}_0, \bm{\mu}_1, \dotsc, \bm{\mu}_{l-1}$) and then enters a constant velocity motion ($\bm{\mu}_l, \bm{\mu}_{l+1}, \dotsc, \bm{\mu}_{l+n-1}$), the scale can only be observed when the states correspond to generic motions are included in the sliding window. However, this will inevitably increase the computation complexity so that the limited computation source of the robot cannot ensure the real-time property. A promising solution is to provide an initial estimate of $\bm{\mu}_l$, then we can propagate the scale from $\bm{\mu}_{l-1}$ to $\bm{\mu}_l$. This can be done by proper marginalization of $\bm{\mu}_{l-1}$ as it is removed from the sliding window at step $l+n$.

Based on the above discussion, we propose a flexible marginalization scheme to address the issue of degenerated motions. Consider a full state vector $\bm{\mathcal{S}} = \left[ \bm{\mu}_0, \dotsc, \bm{\mu}_{n-1} | b_\mathcal{L}\right]$ where $b_\mathcal{L}$ denotes the set of all observed backscatters in the sliding window. We add a state with a new AoA observation $\bm{\mu}_n$ to the sliding window if any of the following three criteria are satisfied: 

\begin{itemize}
	\item The time between two AoAs $\Delta t$ is larger than $\delta$.
	\item The observed backscatter tags change in the new state.
	\item The newest AoA observation significantly differs from the second newest observation in the sliding window.
\end{itemize}

The first criterion aims to bound the error in the integrated result of IMU measurements between two AoAs. Through some tests, we empirically set $\delta$ to be $500$ ms. The second criterion indicates that the system observes new tags that are needed to be localized. The third criterion aims to ensure that the translation of the robot with respect to the observed AoAs is significant. 

To quantify the difference of AoA observations, we define the {\em similarity} between two AoA observations. At each timestamp, an AoA observation is a set of AoAs from multiple backscatters. For the AoA $\theta_i^j$ of $i$\spth backscatter observed at timestamp $j$, we have its direction vector $\mathbf{r}_i^j = [\cos(\theta_i^j), \sin(\theta_i^j)]^\top$. Then we define the AoA observation at timestamp $j$ as
\begin{equation}
	\mathbf{O}^j = [\mathbf{r}_1^j, \mathbf{r}_2^j, \dotsc, \mathbf{r}_{m}^j], \quad \mathbf{O}^j \in \mathbb{R}^{2\times m},
\end{equation}
where $m$ is the number of observed backscatters at time $j$. For any timestamp $k > j$ that the observed backscatters remain unchanged, the similarity can be defined as
\begin{equation}
	\mathcal{M}_{jk} = 1 - \frac{1}{m}\sum_{i = 1}^m \left( \mathbf{O}^j(i)^\top \cdot \mathbf{O}^k(i) \right),
\end{equation}
where $\mathbf{O}^j(i)$ denotes $i$\spth column of $\mathbf{O}^j$. The similarity $\mathcal{M} \in [0, 2]$. The smaller $\mathcal{M}$ the more similar AoA observations. Through experiments, we empirically set a threshold $\varepsilon$. When the similarity of the most recent two AoA observations is larger than $\varepsilon$, it satisfies the third criterion.

The pseudo code is shown in Algorithm~\ref{alg:marginalization}. The algorithm requires that all newly added AoAs $b_{\mathcal{L}^{+}}$ to have at least two observations to succeed in recovering the scale via triangulation (Line 1). Then we set a variable $f = \text{LIFO}/\text{FIFO}$ to indicate whether the system marginalizes out the second newest state $\bm{\mu}_{n-1}$ or the oldest one $\bm{\mu}_0$. The value of $f$ is determined based on whether the new state observed a new tag or the similarity $\mathcal{M}_{n-1}^n$ between two most recent AoA observations (Lines 5--9 and Lines 14--18).

\begin{figure}[t!]
  \centering
  \includegraphics[width=3in]{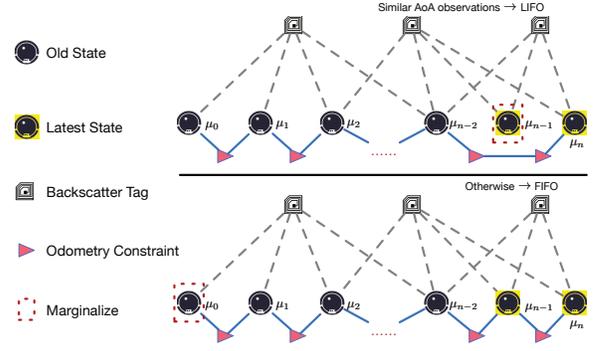}
  \setlength{\belowcaptionskip}{-16pt}
  \caption{An illustration of the flexible marginalization. If the second latest state has a similar AoA observation to the latest one, we will simply marginalize it and all its corresponding AoA measurements. However, pre-integrated odometry measurements are kept and the pre-integration process is continued towards the next state. Otherwise, we will keep it in the window and marginalize the oldest state and its corresponding AoA and odometry measurements. The information of marginalized states is turned into a prior.}
  \label{fig:marginalization}
\end{figure}

To marginalize a chosen state, we first construct a new prior based on all measurements related to the removed state (Lines 3 and 11). We then remove the state, the corresponding AoA observation, and the backscatter tags $b_{\mathcal{L}^{-}}$ that are first observed by it (Lines 4 and 12). The new prior can be expressed as
\begin{equation}
\begin{aligned}
	\bm{\Gamma}_{p}^{+} & = \bm{\Gamma}_{p} + \\
	& \sum\limits_{(i, j)\in \mathcal{A}^{-}} \left(\mathbf{Q}_i^j\right)^\top \left(\bm{\Omega}_i^j\right)^{-1} \mathbf{Q}_i^j + \sum\limits_{k\in \mathcal{I}^{-}} \left(\mathbf{P}_{k+1}^k\right)^\top \left(\bm{\Lambda}_{k+1}^k\right)^{-1} \mathbf{P}_{k+1}^k, 
\end{aligned}
\label{eqn:prior}
\end{equation}
where $\mathcal{A}^{-}$ and $\mathcal{I}^{-}$ are the sets of removed AoA and IMU measurements respectively. The marginalization can be carried out via Schur Complement~\cite{shen2016initialization}. The prior $\bm{\Gamma}_{p}$ is the initial condition computed by solving system~\eqref{eqn:cost}. Eqn.~\eqref{eqn:prior} converts the sum of the Mahalanobis norm corresponding to the removed measurements into a new prior. Note that in the LIFO scheme we have an additional operation that concatenates the IMU odometry from $\mu_{n-2}$ to $\mu_n$ for preserving additional motional information (Line 13).

This approach tries to preserve all information provided by the marginalized states. On one hand, our approach keeps removing the most recent state if the robot has small motion or is stationary. Keeping older states in this case can preserve the non-zero acceleration information that helps recover the scale. On the other hand, when the robot undergoes a constant velocity motion, older states will be removed and the priors implicitly propagate the scale information forward for subsequent estimates. Fig.~\ref{fig:marginalization} illustrates the two working cases of the flexible marginalization approach. In this way, our system needs to incorporate the prior information as follows:
\begin{equation}
	\bm{\mathcal{S}}^* = \argmin_{\bm{\mathcal{S}}} \Big\{\overbrace{\left(\mathbf{b}_p - \bm{\Gamma}_{p}\bm{\mathcal{S}}\right)}^\text{Prior} + \mathbf{A}(\bm{\mathcal{S}}) + \mathbf{D}(\bm{\mathcal{S}}) \Big\}, 
  	\label{eqn:final_cost}
\end{equation}
where $\left\{\mathbf{b}_p, \bm{\Gamma}_p\right\}$ is the prior for our system. The system is then solved with all available measurements within the sliding window plus any available prior (Line 20).

\begin{algorithm} 
\caption{Flexible Marginalization}
\label{alg:marginalization}
\begin{algorithmic}[1]
  \REQUIRE
  				$$\bm{\mathcal{S}} \leftarrow \left[\bm{\mu}_0, \dotsc, \bm{\mu}_{n-1} | b_{\mathcal{L}}\right]$$
  				$$f = \text{FIFO} \; \text{or} \; \text{LIFO}$$
  				$$\left\{\mathbf{b}_p, \bm{\Gamma}_p\right\} \leftarrow \text{Prior}$$
  \ENSURE $\Delta t > \delta$ \OR $b_\mathcal{L}$ changes \OR $\mathcal{M}_{n-1}^n > \varepsilon$
  \STATE $\bm{\mathcal{S}} \leftarrow \bm{\mathcal{S}}\cup\left[\bm{\mu}_n | b_{\mathcal{L}^{+}}\right]$
  \IF {$f = \text{FIFO}$} 
  	\STATE $\left\{\mathbf{b}_p, \bm{\Gamma}_p, b_{\mathcal{L}^{-}}\right\} \leftarrow$ Marginalization$\left(\bm{\mu}_0\right)$
  	\STATE $\bm{\mathcal{S}} \leftarrow \bm{\mathcal{S}}\backslash\left[\bm{\mu}_0 | b_{\mathcal{L}^{-}}\right]$
  	\IF {$\mathcal{M}_{n-1}^n > \varepsilon$ \OR $b_{\mathcal{L}}$ changes}
  		\STATE $f \leftarrow \text{FIFO}$
  	\ELSE
  		\STATE $f \leftarrow \text{LIFO}$
  	\ENDIF
  \ELSE
  	\STATE $\left\{\mathbf{b}_p, \bm{\Gamma}_p, b_{\mathcal{L}^{-}}\right\} \leftarrow$ Marginalization$\left(\bm{\mu}_{n-1}\right)$
  	\STATE $\bm{\mathcal{S}} \leftarrow \bm{\mathcal{S}}\backslash\left[\bm{\mu}_{n-1} | b_{\mathcal{L}^{-}}\right]$
  	\STATE OdometryConcatenation$\left(\bm{\mu}_{n-2}, \bm{\mu}_n\right)$
  	\IF {$\mathcal{M}_{n-2}^n > \varepsilon$ \OR $b_{\mathcal{L}}$ changes}
  		\STATE $f \leftarrow \text{FIFO}$
  	\ELSE
  		\STATE $f \leftarrow \text{LIFO}$
  	\ENDIF
  \ENDIF
  \STATE Solve $\bm{\mathcal{S}}$ using \eqref{eqn:final_cost} and \eqref{eqn:cost_specific} with $\left\{\mathbf{b}_p, \bm{\Gamma}_p\right\}$
  \RETURN $\left\{f, \mathbf{b}_p, \bm{\Gamma}_p, \bm{\mathcal{S}}\right\}$
\end{algorithmic}
\end{algorithm}

\section{Implementation and Evaluation}
\label{sec:evaluation}
In this section, we first present the implementation of Rover based on a iRobot Create 2 and customized backscatter tags. Then we evaluate the performance of individual components as well as the whole system to demonstrate the effectiveness of Rover. 

\subsection{Implementation and Experimental Setup}
\begin{figure}
	\centering
	\begin{minipage}[b]{0.48\textwidth}\centering
		\center
		\includegraphics[width=1\textwidth]{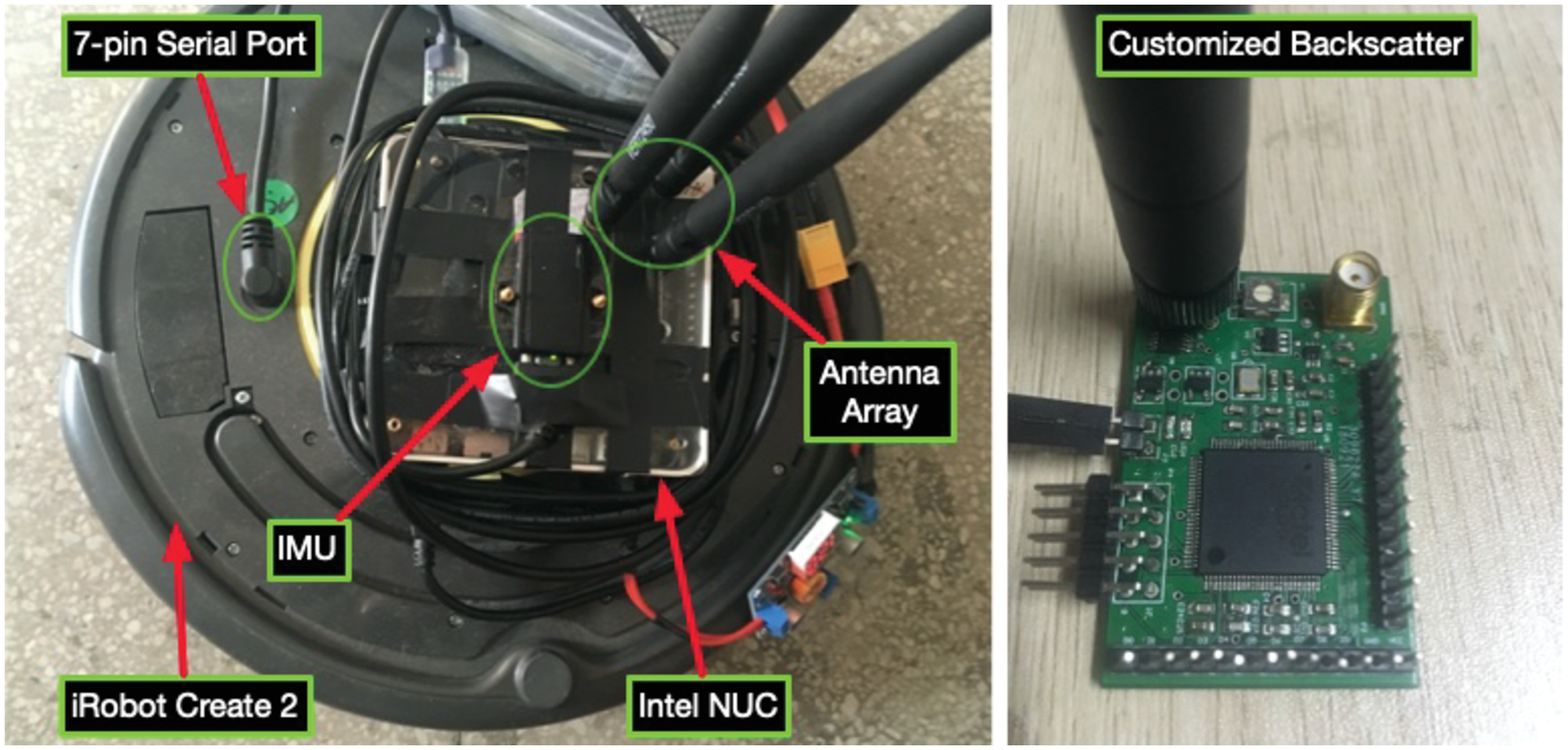}\vspace{-0.3cm}
		\caption{The experimental platform. The left shows the receiver attaches on the robot and sends commands to control its motions through the Create's 7-pin serial port. The right shows one of our customized backscatter tags.} \label{fig:sys}
	\end{minipage}
	\hspace{0.1cm}
	\begin{minipage}[b]{0.48\textwidth}\centering
		\center
		\includegraphics[width=1\textwidth]{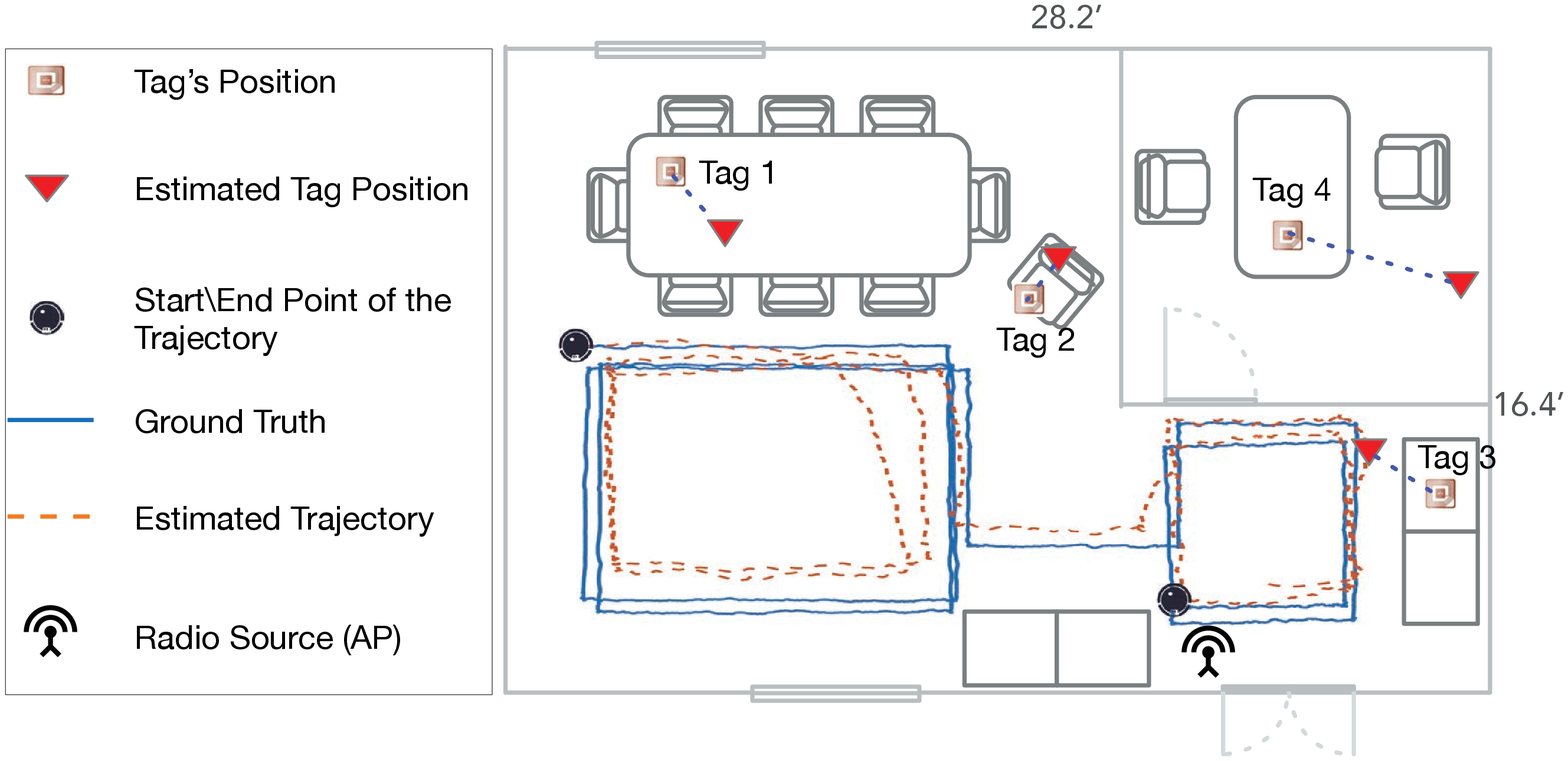}\vspace{-0.3cm}
		\caption{We use the NUC to control a robot moves in a pre-defined rectangular trajectory in the meeting room. The ground truth is provided by the program of defining the trajectory that runs in the NUC.} \label{fig:trajectory}
	\end{minipage}
\end{figure}

We implemented Rover on an Intel NUC with a 1.3 GHz Core i5 processor with $4$ cores, an $8$ GB of RAM and a $120$ GB SSD, running Ubuntu Linux equipped with Intel 5300 NICs and a LORD MicroStrain 3DM-GX4-45 IMU. We use the Linux 802.11 CSI tool~\cite{halperin2011tool} to obtain the wireless channel information for each packet. Thanks to the open-source hardware of HitchHike~\cite{zhang2016hitchhike}, we build the customized tags to backscatter commodity WiFi signals. The power consumption of the tags is only $33\mu$W, $1000\times$ lower than the mW-level power consumption of commodity WiFi. The whole system is implemented in C++. The NUC connects to the iRobot Create 2 and uses ROS (Robot Operating System) as the interfacing robotics middleware to control the robot's moving trajectory\footnote{ROS driver for iRobot Create 2, \url{https://github.com/autonomylab/create_autonomy}.}. The experimental platform is shown in Fig.~\ref{fig:sys}.

In all experiments, we use two NUCs. One is the excitation source that operates in $5.825$ GHz center frequency (channel $165$) on a $20$ MHz band. The other is the receiver on the robot that performs the frequency hopping protocol to sweep all available channels in the $5$ GHz band except channel $165$. 

The experiments are conducted in a $9\times 5$ square meters meeting room in our laboratory, which is a typical indoor setting. Four backscatter tags are deployed in the room. Each tag is configured to shift a frequency and backscatter signals in a separate channel. This prevents the interference between tags. In addition, the frequency shift can be an identifier to distinguish the received signal from which tag as each tag occupies a separate channel.

\subsection{Micro-benchmark Evaluation}

\begin{figure}
	\centering
	\begin{minipage}[b]{0.23\textwidth}\centering
		\center
		\includegraphics[width=1\textwidth]{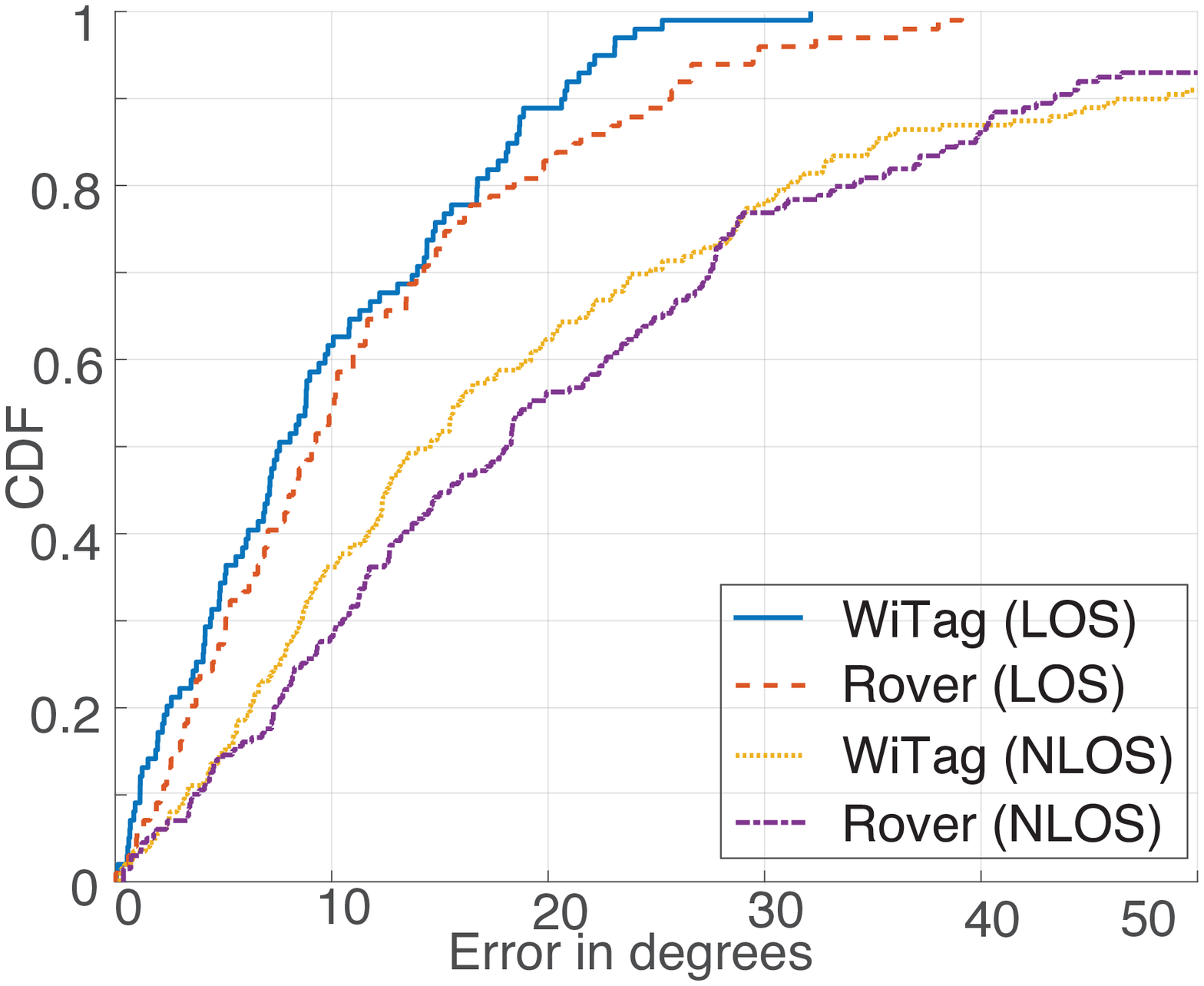}\vspace{-0.3cm}
		\caption{The accuracy of AoA estimation.} \label{fig:aoa}
	\end{minipage}
	\hspace{0.1cm}
	\begin{minipage}[b]{0.23\textwidth}\centering
		\center
		\includegraphics[width=1\textwidth]{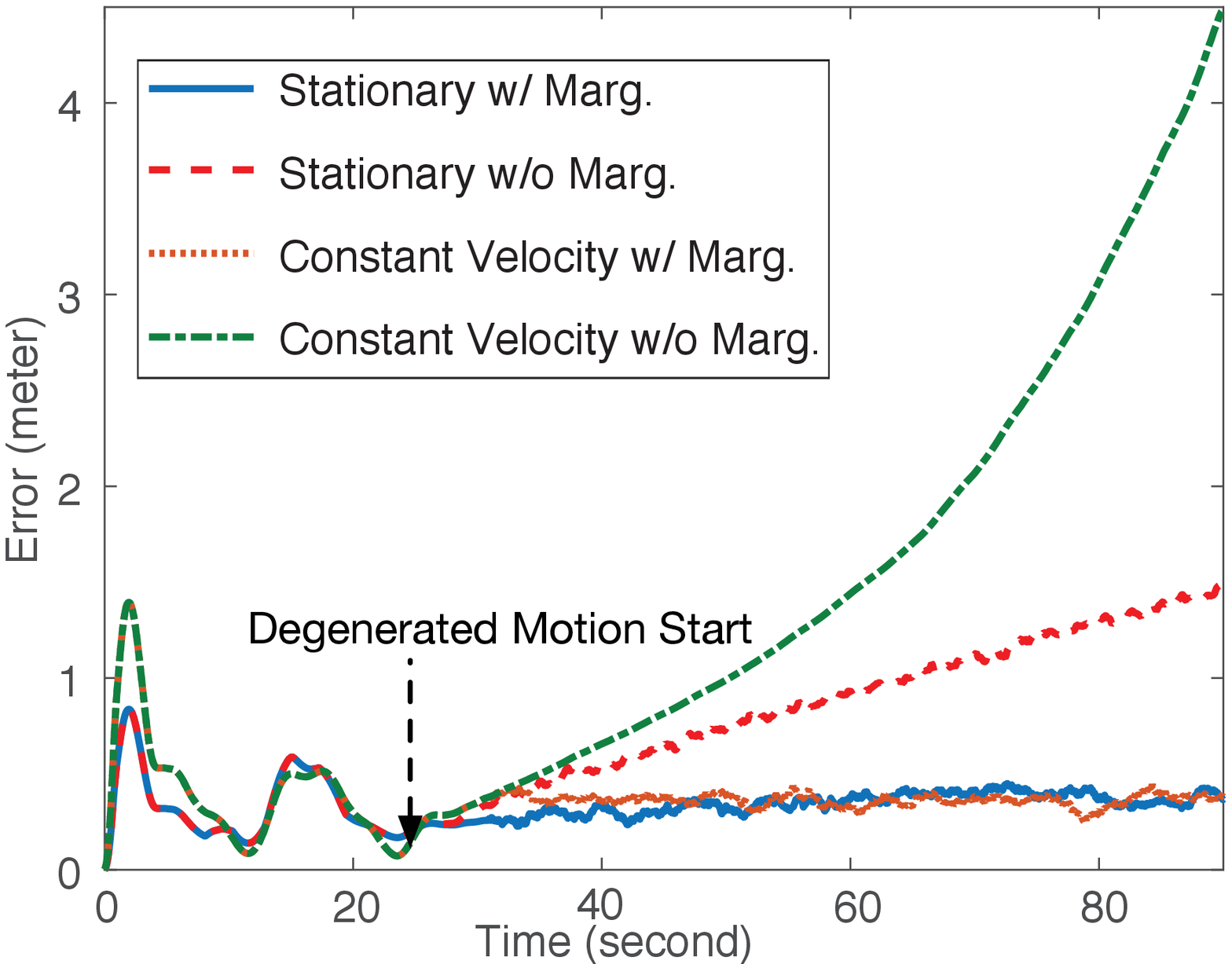}\vspace{-0.3cm}
		\caption{The performance of robot localization under degenerated motions.} \label{fig:degerated_motions_a}
	\end{minipage}
\end{figure}

\begin{table}[t!]
	\centering  
	\caption{Computation time for each update.}	
	\label{table:time}  
		\begin{tabular}{|p{25pt}|p{80pt}|p{80pt}|}  
		\hline  		
		No. of States&Robot Position Accuracy (cm)&Mean Computation Time (ms) \\ 
		\hline
		20&70.2&18.29 \\
		\hline
		30&45.3&27.40 \\
		\hline
		40&38.6&39.21 \\
		\hline
		50&36.5&58.50 \\
		\hline
		60&37.2&99.38 \\
		\hline
		70&37.0&158.17 \\
		\hline
		80&35.9&235.42 \\
		\hline
	\end{tabular}
\end{table}

{\bf Backscatter AoA Estimation}. We first test the accuracy of tag-to-receiver AoA estimation. The key difference in AoA estimation from the state-of-the-art~\cite{kotaru2017localizing}, WiTag, is that we empower it with time-division multiplexing so that the receiver can simultaneously measure AoAs of multiple tags who backscatter signals in different channels. We demonstrate the AoA estimation by four tags deployed in line-of-sight (LOS) and non-LOS (NLOS) settings. The CDF plotted in Fig.~\ref{fig:aoa} shows that the performance of Rover is similar to WiTag. The median errors of Rover and WiTag are $9.3\degree$ and $8.1\degree$ respectively in LOS deployment. In NLOS deployment, the median errors of Rover and WiTag are $18.1\degree$ and $14.6\degree$, respectively.

{\bf Performance under degenerated motions}. We then test Rover's tracking performance under degenerated motions, including being stationary and moving at a constant velocity. Meanwhile, we run Rover in two concurrent processes, with and without marginalization, for the evaluation of our marginalization algorithm. The robot first undergoes generic motions in both tests. Then at $24$ second, it stays stationary in the first test and moves at a constant velocity of $0.1$ m/s in the second test. Fig.~\ref{fig:degerated_motions_a} shows that the localization error accumulates in both cases of degenerated motions if there is no marginalization, but exhibits no accumulation when applied. The robot's mean localization errors are $35.5$ cm and $33.7$ cm during the first $24$ seconds of generic motions. Then the errors go up to $82.1$ cm and $148.7$ cm when being stationary and moving at a constant velocity in the case of no marginalization. When applying marginalization, the errors reduce to $39.1$ cm and $37.0$ cm in the two tests. We notice that the error accumulates faster in the constant velocity movement when marginalization is absent. This is due to the fact that near-zero linear acceleration in this case makes the moving distance unobservable from IMU measurements. Meanwhile, the AoA estimation still changes according to the movements, yielding erroneous results in the SLAM framework. In contrast, when being stationary, the measurements from the AoA and IMU do not contradict each other.

\begin{figure}
	\centering
	\begin{minipage}[b]{0.25\textwidth}\centering
		\center
		\includegraphics[width=1\textwidth]{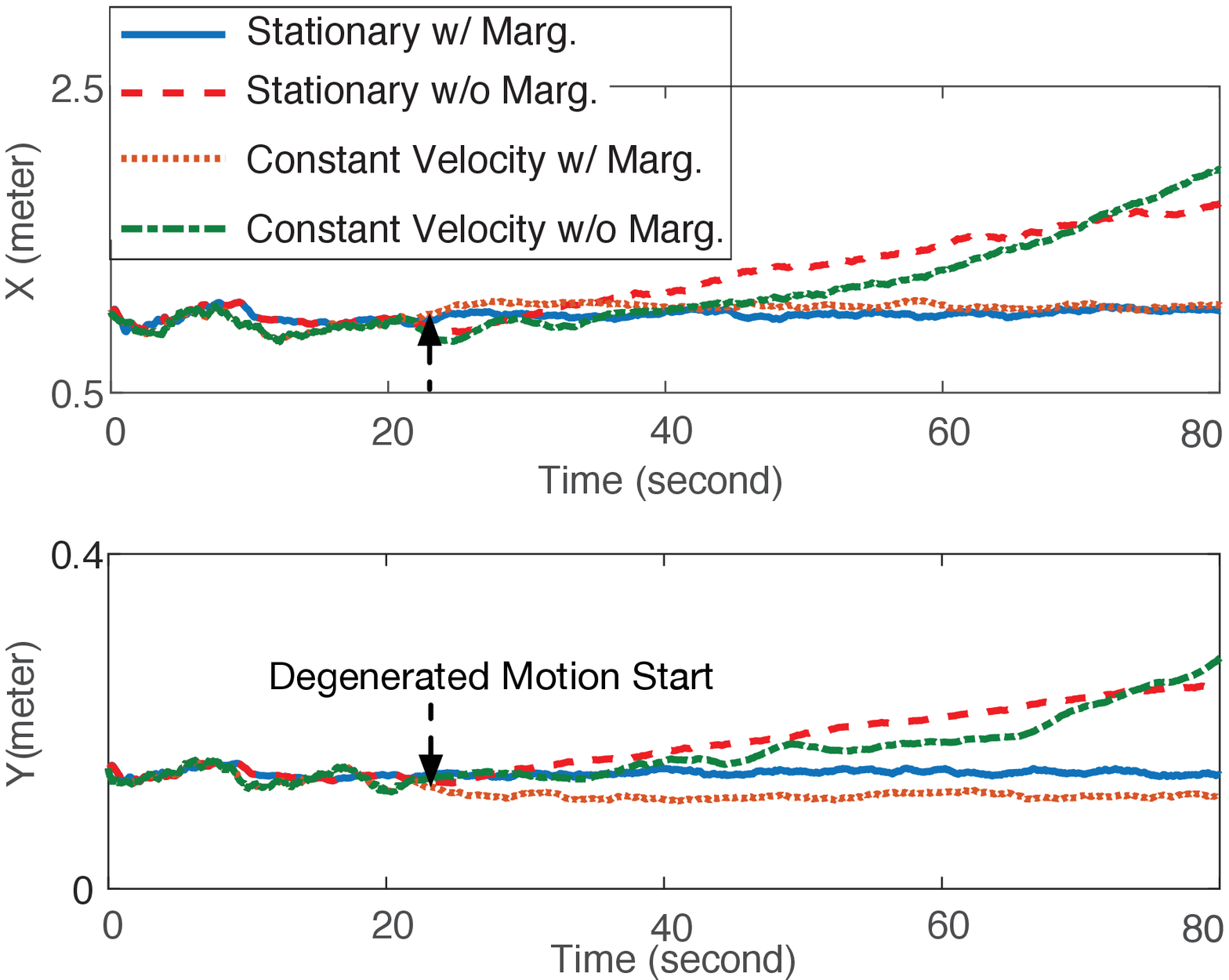}\vspace{-0.3cm}
		\caption{The performance of tag localization under degenerated motions.} \label{fig:degerated_motions_b}
	\end{minipage}
	\hspace{0.1cm}
	\begin{minipage}[b]{0.21\textwidth}\centering
		\center
		\includegraphics[width=1\textwidth]{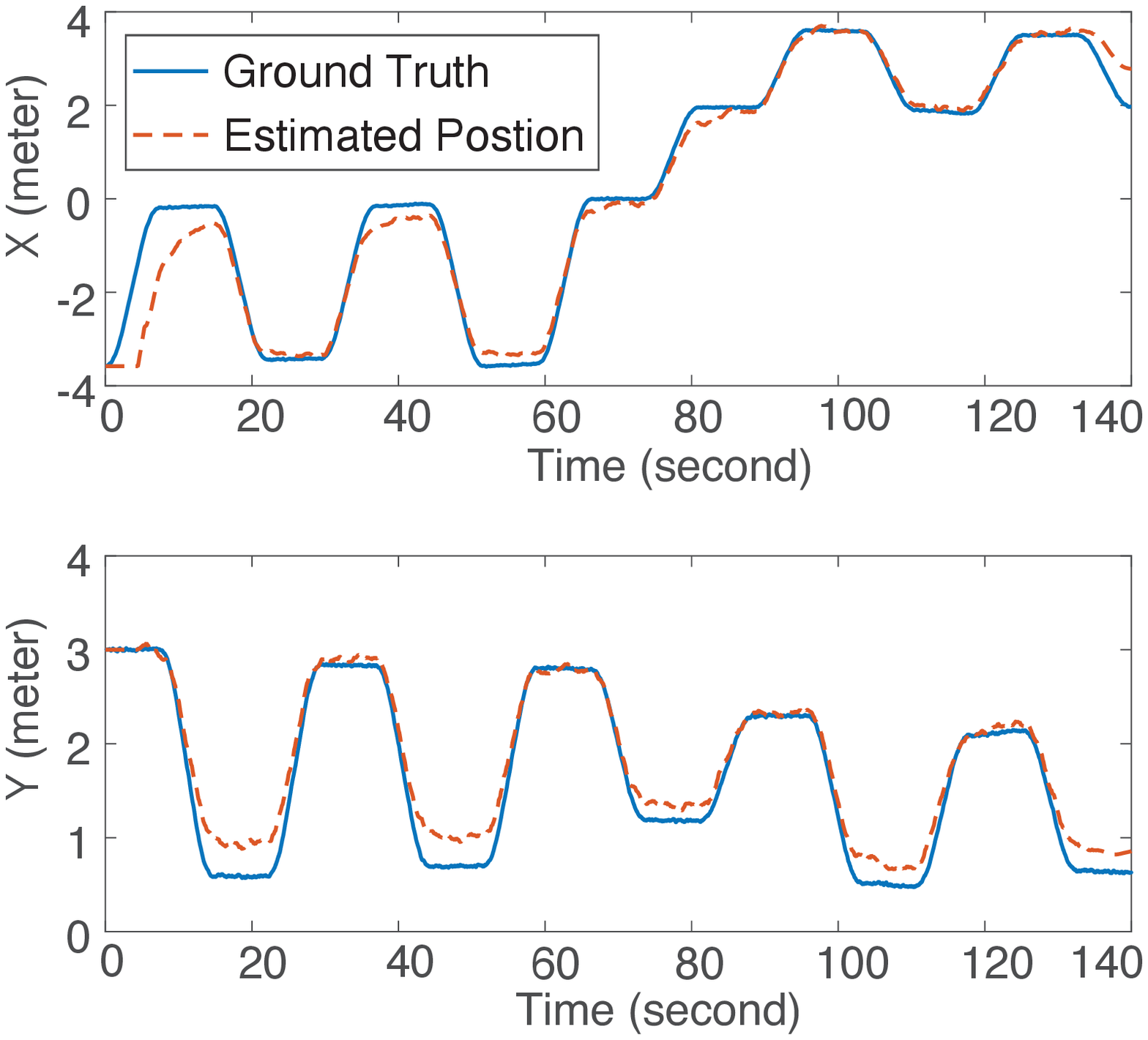}\vspace{-0.3cm}
		\caption{Robot position tracking in the meeting room.} \label{fig:experiment_result_a}
	\end{minipage}
	\vspace{-0.3cm}
\end{figure}

Fig.~\ref{fig:degerated_motions_b} depicts the localization error of a backscatter tag in LOS deployment in different degenerated motions, with and without marginalization. The mean errors of the two tests under generic motions are $76.5$ cm and $83.4$ cm. When the marginalization is not applied, the final errors rise to $118.7$ cm in stationariness and $121.0$ cm in constant velocity motion, respectively. In contrast, applying our marginalization algorithm eliminates the error accumulation. The errors remain $71.3$ cm and $79.6$ cm respectively, which are similar to the performance under generic motions. Again, the error in constant velocity motion accumulates faster due to the erroneous computation. 

{\bf Rover's complexity}. The real-time processing is a desired property in that the real-time location estimates can be used for navigating the robot. Thus, the computation complexity analysis is required. The most time-consuming part of Rover is the SLAM framework, which is a linear system that can be computed quite efficiently. Specifically, we use the standard Cholesky decomposition implemented by Eigen to solve the linear system. The time complexity is $O(N^3)$ in theory, where $N$ denotes the number of states. In practice, the multithreaded routines make the computation time be approximate $N^2$ growth. Despite the mild time complexity, we further employs a sliding window formulation to ensure the real-time processing by bounding the parameter $N$. This is because $N$, which means the number of states in Rover, can be vast in a long-term run and thus significantly increases the computational cost if we solve the full batch SLAM for the best possible accuracy. It poses a tradeoff between localization accuracy and computation time. Essentially, the more states involved the more accurate results obtained. But this inevitably results in higher delay since a larger state vector and the corresponding measurements are involved in the optimization framework. To shed light on that, we tune the number of states in the sliding window from $20$ to $80$ to seek a balance between accuracy and computational cost. Table~\ref{table:time} lists the results in different amounts of states considered. When incorporating more than $50$ states, we can see a marginal increase of the accuracy and a significant increase of the computation time, which goes up to hundreds of milliseconds. Therefore, in our experiments, we empirically set the size of sliding window to be $50$. The overall average computation time is $58.50$ ms for each update and thus Rover achieves the real-time processing.
	
\subsection{System-level Evaluation}

\begin{figure*}
	\centering
	\begin{minipage}[b]{0.3\textwidth}\centering
		\center
		\includegraphics[width=1\textwidth]{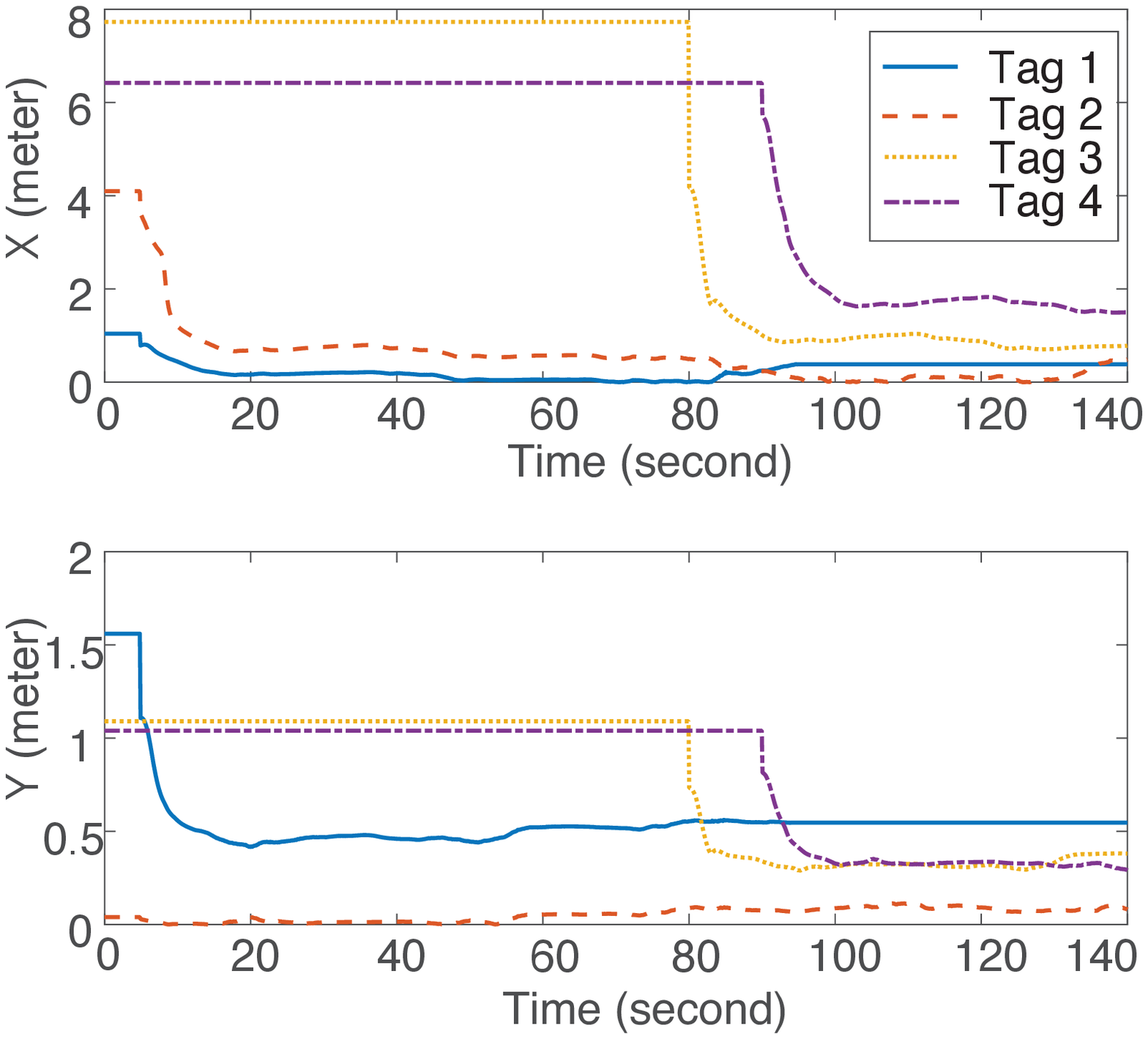}\vspace{-0.3cm}
		\caption{The localization errors of four tags in the meeting room.} \label{fig:experiment_result_b}
	\end{minipage}
	\hspace{0.1cm}
	\begin{minipage}[b]{0.32\textwidth}\centering
		\center
		\includegraphics[width=1\textwidth]{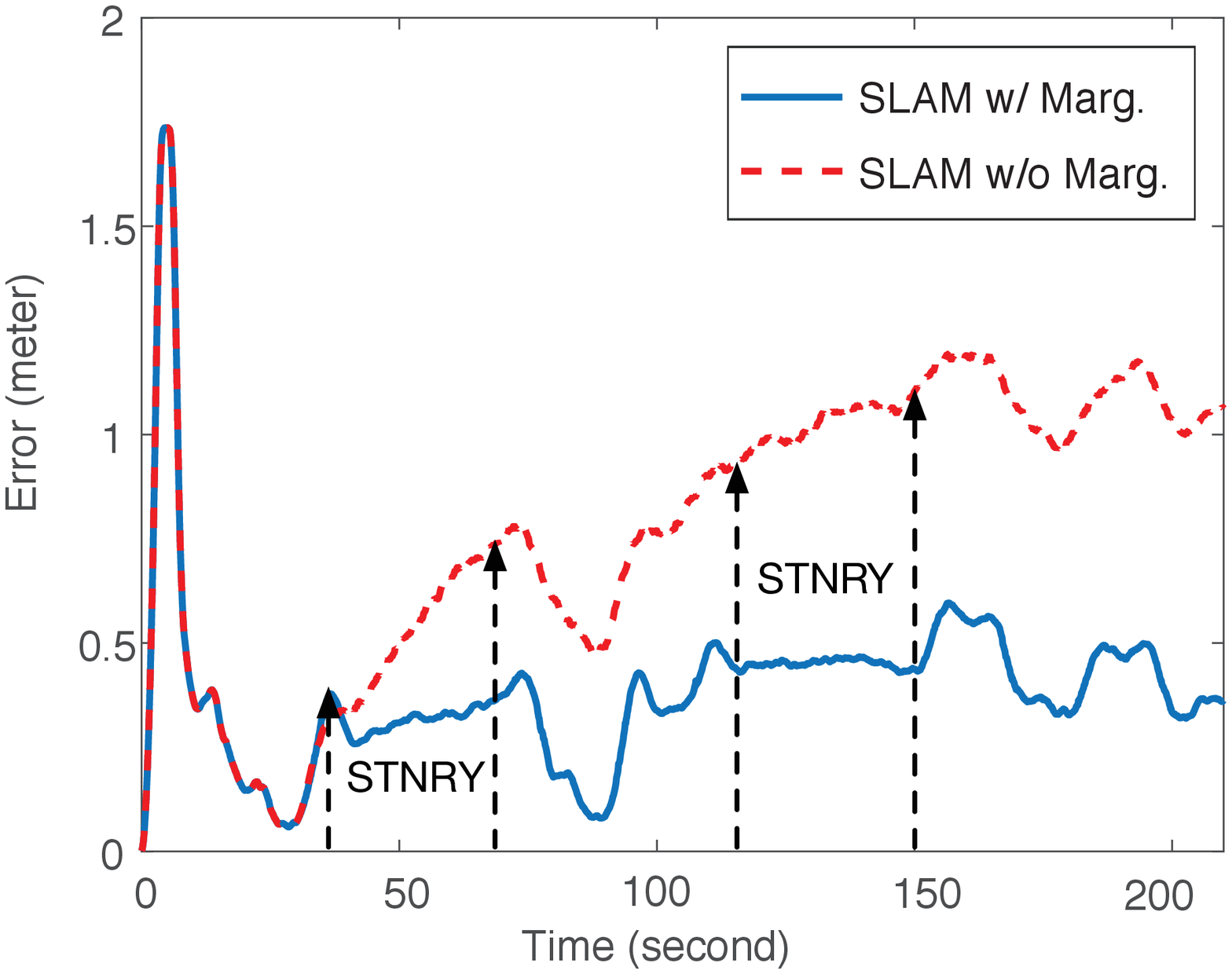}\vspace{-0.3cm}
		\caption{Robot position tracking with degenerated motions. {\bf STNRY} represents {\bf stationary}.} \label{fig:mixed_motions_a}
	\end{minipage}
	\hspace{0.1cm}
	\begin{minipage}[b]{0.32\textwidth}\centering
		\center
		\includegraphics[width=1\textwidth]{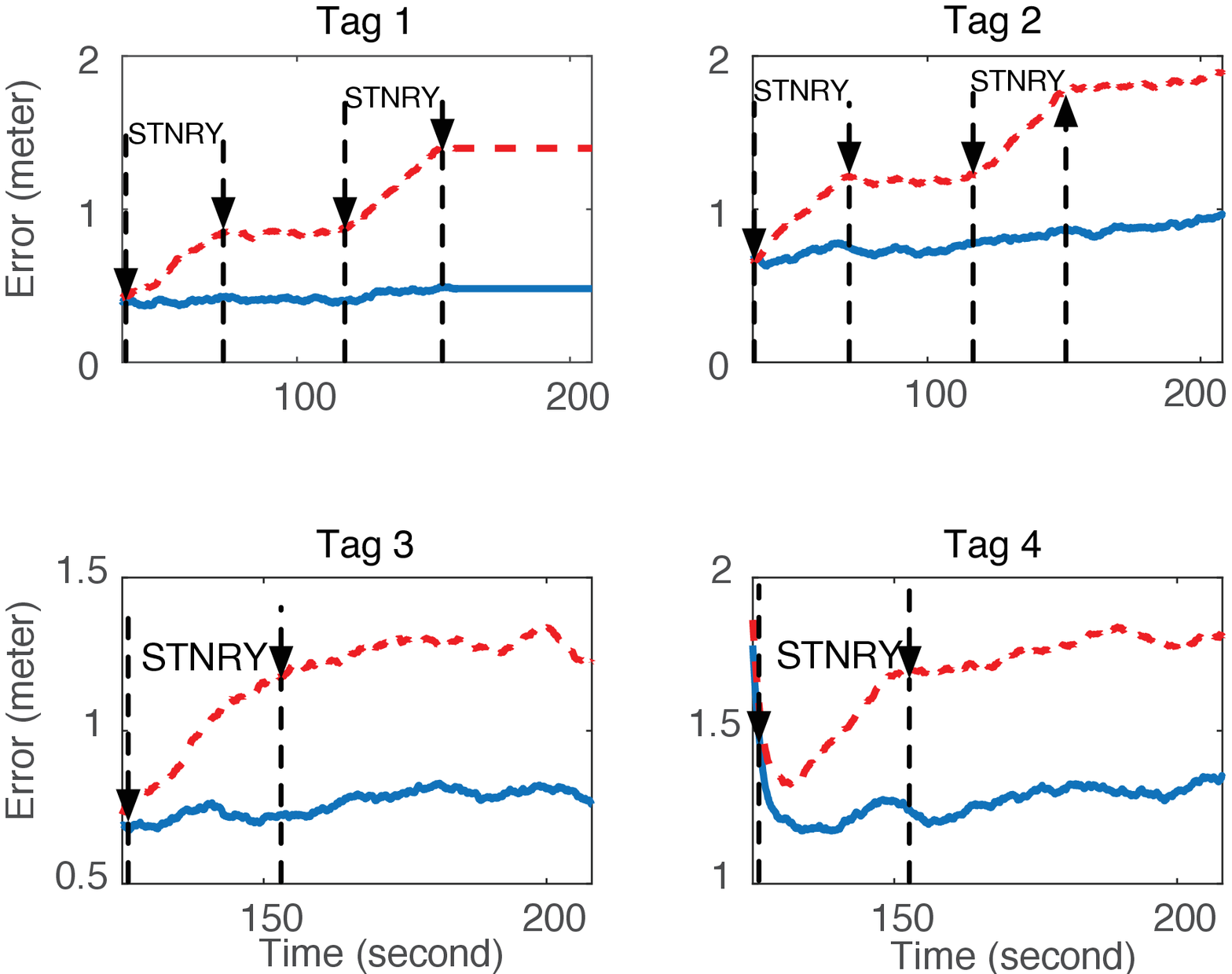}\vspace{-0.3cm}
		\caption{The error reports of four tags' positioning. They are zoomed in to the stationary periods for a better comparison. {\bf STNRY} represents {\bf stationary}.} \label{fig:mixed_motions_b}
	\end{minipage}
\end{figure*}

{\bf Generic motions}. Fig.~\ref{fig:trajectory} shows the system deployment and the overall performance of Rover. The performance of tracking the robot's trajectory is plotted in Fig.~\ref{fig:experiment_result_a}. The mean error over the estimated trajectory is $39.3$ cm. The accuracy goes beyond the expectation from the noisy AoA estimation (Fig.~\ref{fig:aoa}). This is because our system introduces the inertial sensors that provide an additional sensing modality for positioning. Moreover, our sliding window optimization filters out the noise of heterogeneous sensors by finding the configuration of positions that best fits different spatial measurement constraints.  

Fig.~\ref{fig:experiment_result_b} shows the localization results of the tags. Initially, the tags' locations are set to be $(0, 0)$. After about $20$ seconds, Rover localizes tags $1$ and $2$ as their AoAs are available. Tags $3$ and $4$ are localized at about $80$ and $90$ seconds later as the robot approaches them and received their backscattered packets. Meanwhile, Rover stops updating the location of tag $1$ after about $95$ seconds as it loses contact with the robot. The four tags' final localization errors are $73.6$ cm, $52.9$ cm, $97.2$ cm, and $145.9$ cm, respectively. Among them, the error of tag $4$ is higher due to its NLOS deployment. The mean localization error in LOS deployment is $74.6$ cm. 

{\bf Mixed with degenerated motions}. To highlight the effectiveness of our marginalization approach in coping with degenerated motions, we control the robot to stop for a while at some points in the original trajectory and concurrently run Rover in two processes, \ie, with and without the marginalization respectively. Fig.~\ref{fig:mixed_motions_a} shows the position tracking results. The robot stops at $36$\spth second and $125$\spth second, both for a period of $30$ seconds. When incorporating marginalization, there is no sign of error accumulation during the stationary periods. The mean error over the whole trajectory is $43.2$ cm. In contrast, without marginalization, the overall mean error rises to $83.4$ cm.

Fig.~\ref{fig:mixed_motions_b} depicts the localization performance of four tags. For a clearer demonstration, we zoomed into the degenerated periods of each tag. For tags $1$ and $2$, they experience two periods of being stationary at $36$\spth second and $125$\spth second. In the absence of the marginalization, the two tags' final localization errors increase to $146.8$ cm and $182.3$ cm. On the other hand, when enabling the marginalization of Rover, the final errors remain almost unchanged that they are still at $46.4$ cm and $79.3$ cm for tags $1$ and $2$. After about $155$ seconds, tag $1$'s location is no longer updated as it loses the contact with the robot. Similar situation appears on tags $3$ and $4$. Their final errors rise to $124.3$ cm and $178.9$ cm without the marginalization, and remain at $78.2$ cm and $129.2$ cm when enabling the marginalization, which are similar to the performance under generic motions

In summary, the localization accuracy is decimeter-level, which is similar to the state-of-art WiFi based localization systems~\cite{kotaru2015spotfi, kotaru2017localizing}. The uniqueness of Rover is that it works without landmarks or any map of the environment, while conventional solutions need multiple APs with known positions. Conventional solutions use more APs to provide redundant positioning measurements and combat the noise of WiFi measurements. On the contrary, we take advantage of IMU and a robot's mobility to enable a new localization paradigm. The inertial measurements play the role of combating the WiFi noise and the drift-free localizability of WiFi helps correct the IMU drift in return. To bound the computation complexity, we employ a sliding window based formulation and incur a marginalization issue under degenerated motions. Our flexible marginalization algorithm succeeds in addressing the issue.

\section{Related Work}
\label{sec:related}
{\bf Backscatter technology}. Backscatter communication technologies have attracted significant attentions in the last few years to enable low-power and long-term communications~\cite{liu2013ambient, kellogg2014wi, hessar2019netscatter, xu2018backscatter, zhang2016hitchhike, zhang2016enabling, peng2018plora, bharadia2015backfi, kellogg2016passive, liu2018backscatter, amato2018rfid, guo2018design}. Hessar~{\em et al.}~\cite{hessar2019netscatter} proposes a wireless protocol for backscatter networks that supports hundreds of concurrent transmissions. These technologies bring the vision of ubiquitous connectivity into reality for the next-generation of IoT.

{\bf IoT localization}. For many IoT applications, finding such IoT devices is crucial for their smart services. Battery-free RFIDs have been used for localizing IoT devices. Wang {\em et al.}~\cite{wang2013dude} exploits multipath to accurately locate RFIDs in indoors. Furthermore, Luo {\em et al.}~\cite{luo20193d} stitches multiple packets to expand the signal bandwidth and improve the ranging accuracy for 3D RFID tracking. But these approaches need to deploy dedicated RFID readers and they suffers from a limited communication range. To address these issues, Kotaru {\em et al.}~\cite{kotaru2017localizing} presents a WiFi backscatter based localization system. It analyzes the phase information when the received WiFi packet traverses two links, the AP-to-tag link and the tag-to-receiver link. Then it measures the AoAs of a tag to receivers for localization.
Despite the high accuracy, all these approaches require to deploy multiple landmarks, \eg, WiFi APs and RFID readers, whose locations are known to enable the localizability. Calibrating these landmarks to obtain their locations is usually labor-intensive and thus has high start-up cost. In contrast, Rover is plug-and-play to localize IoT devices with backscatter tags without any landmarks or environment knowledge.

{\bf Visual/laser SLAM}. The core of Rover is a SLAM framework that simultaneously localizes the robot and the connected tags via the backscatter RF sensing modality. SLAM has been a long studied problem in the robotic community. Monocular visual SLAM approaches~\cite{fuentes2015visual, huang2017visual} leverage visual data from a monocular camera to localize the robot who equips with the camera and the map represented by point cloud. Shen {\em et al.}~\cite{lin2018autonomous} fuses monocular visual data with IMU measurements to estimate the metric scale of the environment. But the visual sensing is very sensitive to lighting conditions and environmental texture. In addition, laser based SLAM frameworks use laser ranging sensors, \eg, LiDAR, to obtain 3D structure of environments~\cite{dube2017online, hess2016real}. The sensors use emitted light so that they works independent of the ambient light. However, the laser pulse is sensitive to LOS interference like fog or smoke. Overall, visual and laser based SLAM approaches use light-based sensing modalities to estimate the depth map of environments, which are highly accurate and reliable but limited by lighting and LOS conditions. In contrast, Rover's SLAM takes backscatter RF signals as a bridge to connect the robot with the surrounding environment. The RF sensing is complementary to the visual/laser sensing as RF signals can propagate in NLOS settings, traversing obstacles like walls and furnitures thanks to their larger wavelength. The environment in our work is represented by the RF map, \ie, the locations of backscatters. Thus, we localize the tags when solving the mapping problem in our SLAM approach.

{\bf RF-based SLAM}. There have been many SLAM systems that work with RF signals, \eg, WiFi and UWB~\cite{zhang2020wifi, gentner2016multipath, wang2017unified, li2019twc, venkatnarayan2019enhancing, li2016csi}. Huang {\em et al.}~\cite{huang2011efficient} takes the received signal strength of WiFi signal to do the SLAM in indoor environments. They use Gaussian Process Latent Variable Model to reduce the high-dimensional fingerprints to latent-space locations. The RSSI is prone to be noisy in indoors so that the accuracies of these approaches are limited. Recently, Venkatnarayan {\em et al.}~\cite{venkatnarayan2019enhancing} leverages the drift-free localizability of WiFi to correct the drift of IMU and localize users. It can accurately track users' trajectories but does not address the mapping problem, \ie, localizing the WiFi APs. Li {\em et al.}~\cite{li2016csi} use the phase difference of WiFi signals from active radios to infer AoAs for localization. But it only works in outdoors without multipath fading. Gentner {\em et al.}~\cite{gentner2016multipath} and Li {\em et al.}~\cite{li2019twc} exploit multipath components in indoor venues to simultaneously localize the user and the multipath reflection points. Their studies provide great inspirations for our work. Compare with them, the fundamental differences of Rover are that 1) we only use a small antenna array to estimate AoAs with low-power backscattered signals for localization; 2) we handle multipath fading in indoors without any assumptions of the multipath layout; 3) we leverage IMU to measure the metric scale of environments and employ a sliding window fashion to formulate the graph-based optimization problem for better localizing backscatter tags with bounded computation complexity. Moreover, our formulation is very insensitive to the initialization point, which is crucial for conventional filter-based approaches, \eg, extended Kalman filter (EFK)~\cite{gentner2016multipath, li2019twc} and we address the marginalization problem under degenerated motions to make Rover more practical.

\section{Conclusion}
\label{sec:conclusion}
We presented Rover, a backscatter localization system with an AoA-IMU SLAM framework. We formulated a sliding window based model that fused inertial measurements with the AoAs of backscatter tags to a robot measured by commodity WiFi to simultaneously estimate the locations of the robot as well as the connected tags. In addition, we addressed the practical issues of Rover, including real-time processing and data marginalization in degenerated motions. We implemented Rover on the iRobot Create 2 platform attached with an Intel NUC and an IMU. The experiments in both LOS and NLOS indoor settings showed that Rover achieves localization accuracy of tens of centimeters for both the robot and the backscatter tags without any prior knowledge of the work space. Extending our system to work with other wireless devices, such as iBeacon, for better accuracy is an important task for future work.

\bibliography{main}
\bibliographystyle{IEEEtran}

\end{document}